\RequirePackage{fix-cm}
\documentclass[twocolumn,epjc3]{svjour3}  
\smartqed   
\RequirePackage{graphicx}
\RequirePackage{color}
\RequirePackage[colorlinks,citecolor=blue,urlcolor=blue,linkcolor=blue]{hyperref}
 
\usepackage{multirow}
 
\usepackage[T1]{fontenc}
\usepackage[latin1]{inputenc}
 
\usepackage{amsmath}
\usepackage{arydshln}
\usepackage{microtype}
\usepackage[switch]{lineno}
 \usepackage{lineno}
 
\journalname{Eur. Phys. J. C}
\begin{document}

\title{Full background decomposition of the CONUS experiment\\}

\author{H.~Bonet\thanksref{addr1}, A.~Bonhomme\thanksref{addr1}, C.~Buck\thanksref{addr1}, K.~F\"ulber\thanksref{addr2}, J.~Hakenm\"uller\thanksref{a, addr1}, J.~Hempfling\thanksref{addr1}, G.~Heusser\thanksref{addr1}, T.~Hugle\thanksref{addr1}, M.~Lindner\thanksref{addr1},  W.~Maneschg\thanksref{addr1}, T.~Rink\thanksref{addr1}, H.~Strecker\thanksref{addr1}, R.~Wink\thanksref{addr2}}

\thankstext{a}{e-mails: janina.hakenmueller@mpi-hd.mpg.de, conus.eb@mpi-hd.mpg.de}

\institute{Max-Planck-Institut f\"ur Kernphysik, Saupfercheckweg 1, 69117 Heidelberg, Germany \label{addr1}
           \and
          Preussen Elektra GmbH, Kernkraftwerk Brokdorf, Osterende, 25576 Brokdorf, Germany \label{addr2}
}

\date{Received: date / Accepted: date}
\maketitle

\abstract{
The CONUS experiment is searching for coherent elastic neutrino nucleus scattering of reactor anti-neutrinos with four low energy threshold point-contact high-purity germanium spectrometers. An excellent background suppression within the region of interest below 1\,keV (ionization energy) is absolutely necessary to enable a signal detection. The collected data also make it possible to set limits on various models regarding beyond the standard model physics. These analyses benefit as well from the low background level of $\sim$10\,d$^{-1}$kg$^{-1}$ below 1\,keV and at higher energies.
The low background level is achieved by employing a compact shell-like shield, that was adapted to the most relevant background sources at the shallow depth location of the experiment: environmental gamma-radiation and muon-induced secondaries.
Overall, the compact CONUS shield including the active anti-coincidence muon-veto reduces the background by more than four orders of magnitude. The remaining background is described with validated Monte Carlo simulations which include the detector response. It is the first time that a full background decomposition in germanium operated at reactor-site has been achieved. Next to the remaining muon-induced background, $^{210}$Pb within the shield and cryostat end caps, cosmogenic activation and air-borne radon are the most relevant background sources. The reactor-correlated background is negligible within the shield.
The validated background model together with the parameterization of the noise are used as input to the likelihood analyses of the various physics cases. 

}

\section{Introduction}

For every rare event experiment an excellent background suppression in the region of interest (ROI) is crucial for its success.
Coherent elastic neutrino nucleus scattering (CE$\nu$NS) is such a kind of elusive interaction \cite{cevnsfreedman}. It was detected for the first time with a CsI detector at a pion decay-at-rest source by the COHERENT collaboration in 2017 \cite{coherentcsi}. The CONUS experiment aims to measure CE$\nu$NS at the commercial nuclear power plant of Brokdorf (KBR), Germany, with four point contact high-purity germanium (HPGe) spectrometers \cite{cevnspub}. The signal induced by the nuclear recoil from the neutrino scattering is expected to be an excess above the background in the energy range below a collectible ionization energy (unit: eV$_{ee}$) of 500\,eV$_{ee}$. Next to a very low detector energy threshold of below $\sim$300\,eV$_{ee}$, the background within this energy range needs to be suppressed as much as possible to be able to detect a signal. Moreover, it is very important to guarantee a stable background rate without any reactor-correlated contributions. Extending the ROI to higher energies, searches for beyond the standard model (BSM) physics such as for the neutrino magnetic moment \cite{nmmpaper}, scalar and vector mediators as well as non-standard interactions in the neutrino-quark sector \cite{bsmpaper} can be carried out. They benefit from a low and stable background as well. All signal searches are looking for an excess above the background continuum with a particular shape related to the respective model.

For the CONUS experiment, the strategy was pursued to avoid as much as possible background intrinsic to the shield and detector materials beforehand, such as radioactive contaminations and cosmic activation. Next, the shield, adapted to the major background sources and given experimental conditions, was set up. High density materials are mandatory to reduce the environmental gamma-ray ($\gamma$-ray) background. However, as the experiment is located under a shallow overburden of effective 24\,meters of water equivalent (m~w.e.) muon-induced ($\mu$-induced) secondaries created in the high-density materials need to be suppressed. The GIOVE \cite{giovepub} detector at 15\,m~w.e., in which background levels comparable to detectors located several hundreds of m~w.e. deep were achieved, served as inspiration for the CONUS shield design in this regard. This includes an active muon rejection system (in the following referred to as "$\mu$-veto") and layers of borated polyethylene (PE) in a compact shield design, which also meets the spatial constraints at the reactor site.
Besides $\mu$-induced neutrons, also any potential fission neutron background from the reactor site needs to be carefully characterized and shielded, as the neutron recoils can potentially mimic CE$\nu$NS signals. This has been successfully achieved by the CONUS shield as described in \cite{neutronpub}.
All in all, a background level of 10\,d$^{-1}$kg$^{-1}$ within the ROI for CE$\nu$NS was obtained.  

The remaining background within the shield is decomposed into the different contributions with the help of Monte Carlo (MC) simulations with the aim to understand and fully reproduce the spectrum. While the physics searches are carried out below 10\,keV$_{ee}$, the measured spectra at higher energy help to constrain the contributions from the different background sources.
Special attention needs to be paid to the time-dependent background contributions (e.g. reactor-correlated background, airborne radon (Rn), decay of cosmogenic induced radioisotopes with a half-life of one year or less), which can potentially create differences in data collected during reactor OFF and ON periods.

The resulting full background description is an input for the likelihood analyses looking for CE$\nu$NS signals \cite{cevnspub} as well as the searches for BSM physics \cite{bsmpaper}. The MC background model enables a distinction between the understood physical background and additional noise contributions towards lower energies within the ROI for CE$\nu$NS at the energy threshold of the detectors. This contribution needs to be described separately to complete the background model.

This article is structured as follows: In Section \ref{chapter2} an overview of the CONUS shield is given, the overall background suppression capability is discussed and the preparation of the data sets for the background model is described.
This is followed by an introduction of the MC simulation framework including all relevant efficiencies in Section \ref{chapter3}. 
The next section (Section \ref{chapter4}) discusses the different background sources and their contribution to the overall background as predicted by MC as well as their variability in time in the following order: $\mu$-induced background, cosmic activation of Ge and copper (Cu), $^{210}$Pb within the innermost shield layer made out of lead (Pb) and the cryostat end caps, airborne Rn, other contaminations within the cryostat, and reactor-correlated background.
Finally, in Section \ref{chapter6} the previously described background components are combined forming the background model, which is used together with the noise contribution as description of the reactor OFF data in the CONUS physics analyses. In the end, an outlook on potential improvements in the background suppression will be given.

\section{Experimental input for the background model}
\label{chapter2}

\subsection{The CONUS experiment and detectors}

The CONUS experiment is located in room A-408 at KBR at a distance of 17.1\,m to the reactor core, which guarantees a high antineutrino flux of 2.3$\cdot$10$^{13}$\,s$^{-1}$cm$^{-2}$. 
The experiment employs four low energy threshold point contact high-purity Ge (HPGe) spectrometers with a mass of 1\,kg each, in the following labeled with C1 to C4. The detectors are described in detail in \cite{detectorpub} and the characteristics relevant for the MC simulations are listed in Section \ref{sec:mcsim}. The detectors have an excellent energy resolution, which translates into a low noise threshold required to access the CE$\nu$NS signal region.

\begin{figure*}
    \centering
    \includegraphics[width=0.85\textwidth]{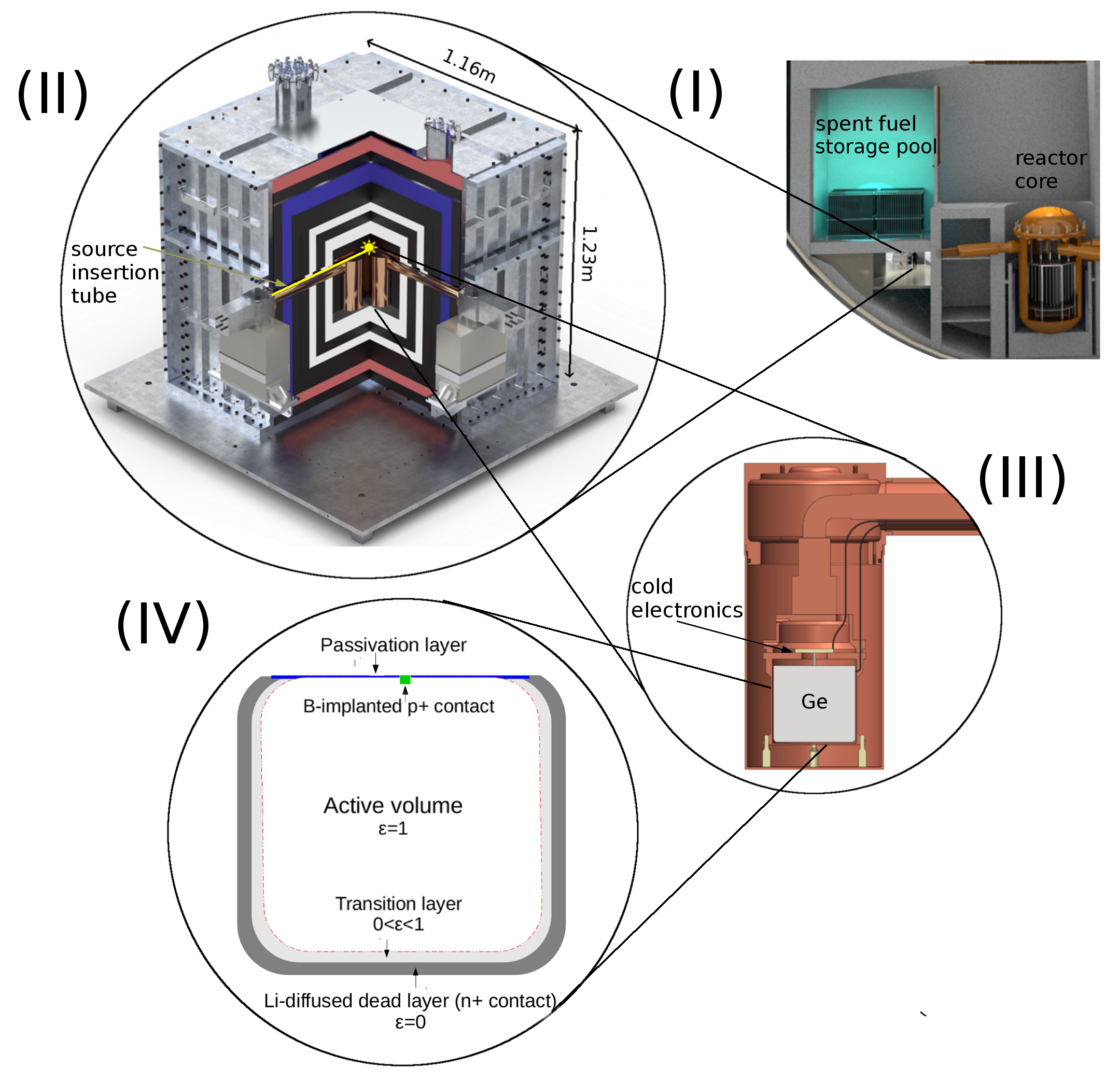}
    \caption{ (I) The CONUS experiment is located below the spent fuel storage pool at KBR inside room A-408. The pool together with the surrounding building provides an effective overburden of 24\,m~w.e. at a distance of 17.1\,m to the reactor core. The compact shield (II) adapted to this location consists of 25\,cm of Pb (black), 10\,cm of borated PE (white) and pure PE on top and bottom (red) as well as the $\mu$-veto, which is made up out of plastic scintillator plates of 6\,cm thickness equipped with PMTs (blue). The shield is enclosed in a steel cage for safety reasons. The Cu cryostats of two of the four detectors are visible in the figure. A PE tube (marked in yellow) for the insertion of calibration sources inside the shield is located above one detector. (III) The detectors themselves are point-contact HPGe spectrometers consisting of a Ge diode in a Cu housing. (IV) Outside of the active volume (charge collection efficiency $\epsilon=1$) of the Ge diode within the transition layer the efficiency is diminished ($\epsilon<1$), and energy depositions within the dead and passivation layer on the surface of the diode are not registered at all ($\epsilon=0$).}
    \label{fig:chapter2_conusshield}
\end{figure*}

At A-408, an overburden of 10-45\,m~w.e. is provided depending on the solid angle to shield against cosmic radiation. Especially the spent fuel storage pool, the pool for the transport casket of the spent fuel elements and the surrounding concrete structures directly above the experimental site suppress the $\mu$-flux and the hadronic component of the cosmic rays. The experimental site is shown in Figure \ref{fig:chapter2_conusshield} and described in detail in \cite{neutronpub}. The shallow overburden means that special attention has to be paid to the suppression of the remaining $\mu$-induced background.

Next to natural radioactivity prevalent everywhere in the surrounding structures, neutrons from the reactor core and high energetic $\gamma$-radiation from the cooling cycle of the primary circuit contribute to the ambient radiation. This reactor-correlated background has to be taken into account in the shield design as well. 

Before the installation of the experiment at the power plant, the detectors and the shield were tested in the low level laboratory (LLL) of the Max-Planck-Institut f\"ur Kernphysik (MPIK), Heidelberg, Germany, with an overburden of 15\,m~w.e. The shield was assembled and disassembled several times as described in the next section with various combinations of detectors within. Among others, the radiopurity of the shield materials was evaluated with the known radiopure coaxial HPGe spectrometer CONRAD (active mass 2.2\,kg) (see \cite{neutronpub}). Its large active mass ensures a reasonable high detection efficiency up to 10\,MeV for $\gamma$-rays, which covers the full range of natural and artificial radioactivity.

\subsection{Data collection}
\label{subsec:dataforbkgmodel}
 
The CONUS shield was constructed in 2017 and thoroughly tested during the commissioning at the LLL in five phases, lasting for several weeks each. 
In the first phase of data collecting, only the CONRAD detector was placed inside the shield to check the radiopurity of the shield materials. The CONRAD background data are discussed in Section \ref{subsec:muons} and Section \ref{subsec:pb}.

In the next three phases, different combinations of the four CONUS spectrometers were enclosed inside the shield. The background data collected in this way can be compared to the data from the nuclear power plant and is included in the study of the decay of the cosmogenic induced background lines (see Section \ref{subsec:muons} and Section \ref{subsubsec:cosmicdata}).  
In the last phase, the neutron fluence rate inside the shield was determined directly with a Bonner Sphere. The results are reported in \cite{mybyby}.

At the beginning of 2018, the CONUS experiment was set up at KBR inside room A-408. For the background studies discussed in this publication, data collected during the onsite commissioning, the physics data collection of RUN-1 (April 1 - October 29, 2018) and RUN-2 (May 16 - September 23, 2019) and the optimization phase in-between are used. These include the reactor outages of both physics runs. At the end of the outage in June 2019 a leakage test was carried out at the power plant. The pressure inside the containment building was increased to 1.5\,bar. To guarantee the safety of the detectors, the cryostats were ventilated and filled with high purity gaseous Argon \cite{detectorpub}. After the leakage test, the vacuum was restored by pumping the detectors. However, due to irreversible changes in the performance and background level of the detectors the data collected afterwards requires adaptions of the background model not included in this publication. The background level below 10\,keV$_{ee}$ increased by up to a factor of two, depending on the detector.

C4 is not included in the analyses of RUN-1 and RUN-2 data due to an artifact observed at lower energies in RUN-1 and incompatibilities in the DAQ settings for RUN-2 reactor ON and OFF exposure before the leakage test. Therefore, the detector was excluded from the background examinations presented here, but overall the background composition is comparable to the other three detectors.
 
The CONUS experiment employs the Lynx data acquisition system (DAQ) (described in \cite{detectorpub}). Five channels are available. As the experiment focuses on the low energy region towards the noise threshold, the standard energy range selected for data collection during RUN-1 and RUN-2 for each detector is below 15\,keV$_{ee}$ (in the following labeled as "lowE") to enable a detailed coverage of this energy range. However, to fully understand and study the background a larger spectral energy range is beneficial. Therefore, for one of the four detectors the signal is split to the fifth available DAQ channel set up to the maximal energy range of $\sim$500\,keV$_{ee}$. This is referred to as "highE" in the following. The upper limit is defined by a combination of the limitations of the Lynx and of the dynamic range of the preamplifier. For RUN-1, C1 was selected, while starting with RUN-2 C3 has been used for the monitoring of the highE range. For C3, due to the enhanced Rn background during the OFF time (see Section \ref{sec:airbornern}) also data from the subsequent reactor ON time after the complete suppression of the Rn background has been used for the construction of the model. For the other detectors only the statistics collected during commissioning or optimization phases is available in the highE range. Consequently, it is not possible to use just reactor OFF data for the background model of all detectors, but reactor ON data also needs to be taken into account. Moreover, within the Lynx DAQ system the $\mu$-veto is applied online during data collection. Data without $\mu$-veto have to be collected separately, which is done sporadically.   

All data from the CONUS detectors are processed applying the time difference distribution cut (TDD cut) as described in \cite{detectorpub} to remove noise and events incorrectly reconstructed by the DAQ. The cut is only relevant for energies below 50\,keV$_{ee}$.
The background model below 15\,keV$_{ee}$ is developed with the same reactor OFF data sets as used for the CE$\nu$NS analyses \cite{cevnspub} and part of the BSM analyses \cite{bsmpaper}. To obtain these data sets additionally the noise-temperature correlation cut (NTC cut), which strongly impacts the exposure, is used as described in \cite{detectorpub}.  

An overview of the data sets considered for the background model is given in Table \ref{tab_bkgmodeldataexposure}. In the following, the region of interest for CE$\nu$NS is set to [0.3,1]\,keV$_{ee}$. To compare MC spectra to data, this range is restricted to [0.4,1]\,keV$_{ee}$ due to the increase of electronics noise in the data below 400\,eV$_{ee}$, which is described by a fit function added to the MC background model. Additionally, a second ROI is defined from [2,8]\,keV$_{ee}$ for the analyses related to some BSM models.

\begin{table}[btp]
\caption{Data sets with $\mu$-veto applied used for the construction of the background model. Regular highE data can only be collected for one detector during a run due to DAQ limitations. From this it follows that reactor ON data need to be used for the construction of the background model for some detectors. ($^{(*)}$data collected in optimization phase after RUN-1) \label{tab_bkgmodeldataexposure}}
\begin{tabular}{llll}
\hline\noalign{\smallskip}
det.&Run &   lowE [kg$\cdot$d]    & highE [kg$\cdot$d] \\  
         &    &   \multicolumn{2}{l}{(reactor status)}\\ \hline          
C1       &  1  &  13.8 (OFF)  &    30.0 (OFF)     \\
         &  2  &  12.1 (OFF)  &         \\
C2       &  1   &  13.4 (OFF)  &  26.6 (ON$^{(*)}$)    \\ 
         &  2  &  9.1 (OFF)  &         \\ 
C3       &  1  &  10.4 (OFF)  &         \\ 
         &  2   &  9.1 (OFF)  & 20.0 (OFF)       \\ 
         &      &       &       26.4 (ON)        \\       

\noalign{\smallskip}\hline

\end{tabular}
 
\end{table}

\subsection{The CONUS shield}
\label{chapter2_shield} 
\begin{figure}[h]
    \centering
    \includegraphics[width=0.45\textwidth]{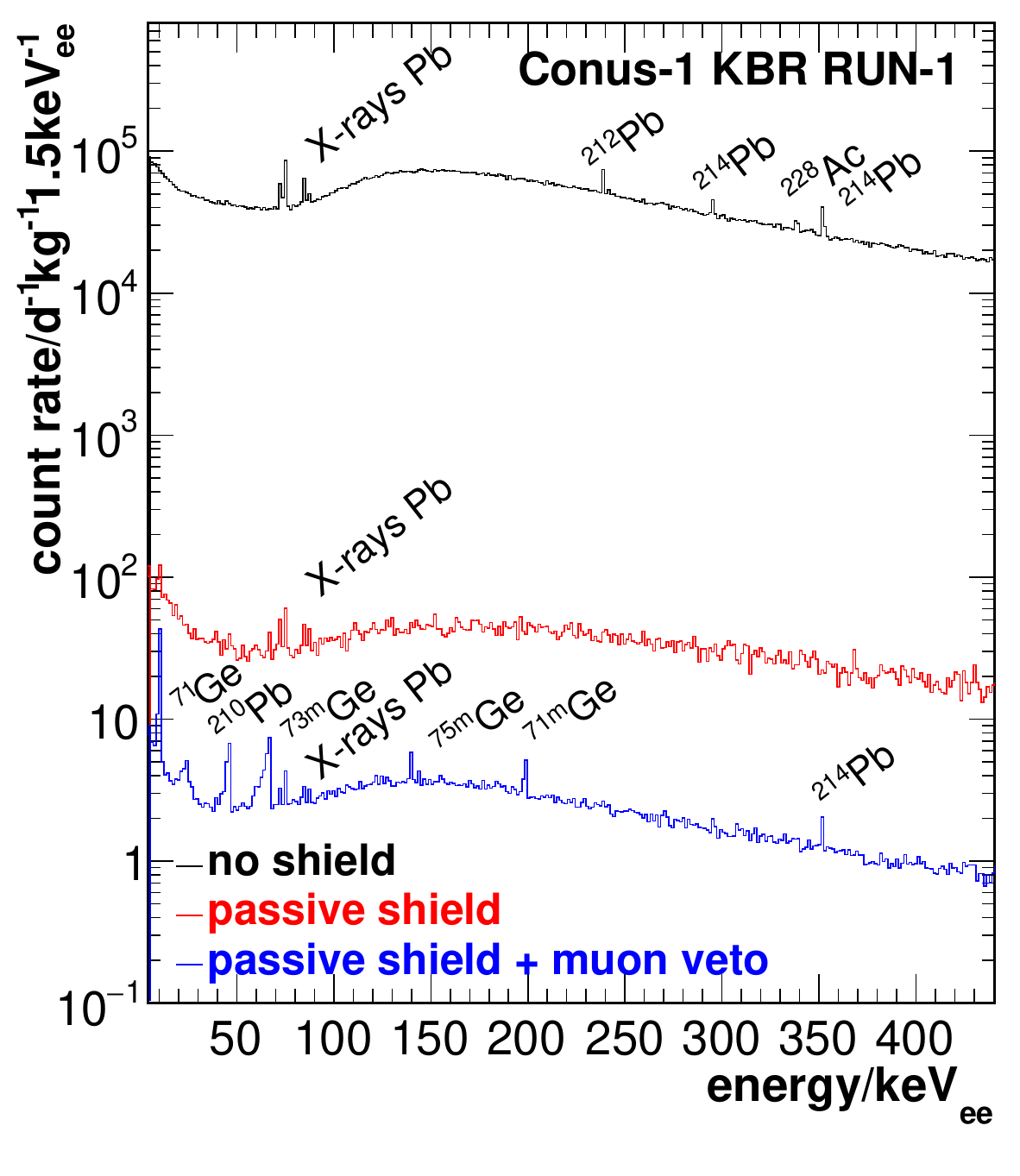}
    \caption{Background suppression capability of the CONUS shield demonstrated for C1: Without any shielding the black spectrum containing background lines of natural radioactivity is detected. The red spectrum is achieved by the passive shield alone. It is dominated by $\mu$-induced background. The blue line corresponds to the background with the $\mu$-veto applied, where background lines, mostly from the decay of metastable states from neutron capture in Ge, become visible.}
    \label{fig:chapter2_bkgsuppression_conusshield}
\end{figure}

In the following, the CONUS shield is introduced and an overview of its background suppression capabilities at KBR is given. 
Without any shield the background induced by the ambient $\gamma$-radiation at KBR dominates over all other background contributions as can be seen in Figure \ref{fig:chapter2_bkgsuppression_conusshield}. The background lines from natural radioactivity are visible. Additional to the background by natural radioactivity, partial energy depositions of higher energetic $\gamma$-radiation from neutron capture and $^{16}$N decays within the cooling cycle of the nuclear power plant contribute to the continuum. 
The $^{16}$N nuclei are produced in (n,p) reactions on oxygen in the reactor core and due to the long enough half-life of 7.13\,s of the isotope the decay occurs within the cooling cycle close to the location of the CONUS experiment \cite{neutronpub}.

The CONUS shield depicted in Figure \ref{fig:chapter2_conusshield} consists of 25\,cm of Pb to suppress this environmental radioactivity. However, muons create secondary electromagnetic radiation and neutrons within the Pb.

In between, layers of borated PE (manufactured with boron (B) enriched in $^{10}$B, 3\% equivalent of natural B) and pure PE (10-15\,cm) are placed to moderate these $\mu$-induced neutrons, but also neutrons from other sources outside the shield. The layer of plastic scintillator plates ($\sim$6\,cm thick) works as moderator as well. 
Up to four HPGe detectors can be placed inside the detector chamber. The cryostats, the supporting structures within, and the cooling fingers are made out of Cu due to the high level of radiopurity of this material. The electrical cryocooler units, which on the contrary are performing poorly in this respect, are located outside of the shield. Consequently, the cooling fingers need to be long enough to pass the full shield of 42\,cm. This requirement put a challenge on the detector design regarding the cold transfer, but it was successfully managed.

The whole shield is enclosed in a steel cage ($\sim$1\,cm thick) that can be flushed with air from compressed air bottles to suppress the airborne Rn background. Freshly filled air had been stored in the bottles for at least two to three weeks, such that $^{222}$Rn (T$_{1/2}$=3.8\,d) decays before the usage of the bottles. 

This makes up the "passive shield", which suppresses the background by two to three orders of magnitude. The spectral shape of the remaining background is governed by the {\it bremsstrahlung} continuum of $\mu$-induced secondaries. The only visible background lines are X-rays from Pb and potentially the 10.37\,keV line of $^{68}$Ge/$^{71}$Ge, if the detector was activated enough before. The 511\,keV line from e$^{+}$e$^{-}$ annihilation is just out of the highE range.

The "active shield" refers to the $\mu$-veto anticoincidence system. The plastic scintillator plates EJ-200 by Scionix are equipped with 2-4\,photomultiplier tubes (PMTs) each (20\,PMTs in total) to detect the muons that cross the CONUS setup. The plastic scintillators are also sensitive to ambient $\gamma$-rays. Therefore, they are located inside the outermost layer of Pb to be able to set a low trigger threshold such as to register most muons without detecting too many $\gamma$-rays. Each signal from the $\mu$-veto trigger opens a constant veto window of 410\,$\mu$s, which is applied online to all Ge data. This results in a dead time of (3.5$\pm$0.2)\,\% for reactor OFF and (5.8$\pm$0.2)\,\% during reactor ON time. The dead time is increased when the reactor is ON due to the enhanced reactor-correlated $\gamma$-ray flux registered by the plastic scintillator plates. The dead-time is precisely monitored and accounted for in the analysis. With the help of the $\mu$-veto, the background can be successfully suppressed by at least one additional order of magnitude as can be seen in Figure \ref{fig:chapter2_bkgsuppression_conusshield}. 
The composition of the remaining background in blue is examined in detail with the help of the MC simulations. Four of the lines observed within the available highE energy range are created by the capture of $\mu$-induced neutrons on Ge. Those metastable states have a half-life longer than the $\mu$-veto window and are therefore nearly not diminished. Moreover, lines from cosmic activation can be seen. The only clearly visible line from contaminations within the cryostat end cap is the 46.5\,keV line from the decay of $^{210}$Pb. As discussed in Section \ref{sec:airbornern}, for certain time periods, the 351.9\,keV line of $^{214}$Pb from the decay chain of airborne Rn might be visible. 

All components within the cryostat end cap (see Section \ref{sec:contaminations}), the borated PE and the innermost layer of Pb (see Section \ref{subsec:pb}) were tested beforehand on potential radioactive contaminations by material screening measurements. 

With the described shield, background levels of 5-15\,d$^{-1}$kg$^{-1}$ in [0.4,1]\,keV$_{ee}$, 25-40\,d$^{-1}$kg$^{-1}$ in [2,8]\,keV$_{ee}$ and 400-800\,d$^{-1}$kg$^{-1}$ in [11,440]\,keV$_{ee}$ were achieved. The detailed count rates for each detector and energy range are listed in Table \ref{tab:bkgmodeltable}. Overall, a background suppression of more than four orders of magnitude could be demonstrated at shallow depth with a compact shield of a mass of $\sim$10.9\,t and a volume of 1.65\,m$^{3}$.

\section{MC simulation framework}
\label{sec:mcsim}
\label{chapter3}

The remaining background after the application of all cuts is decomposed by MC simulations. The simulations are carried out with the framework MaGe \cite{mage}, based on Geant4.10.3p03 (\cite{geant4_1}, \cite{geant4_2}). 
MaGe is well validated for low energetic electromagnetic processes (e.g \cite{giovepub}, \cite{budjas2009}, \cite{Poon2005}, \cite{Hurtado2004} and \cite{Amako2015}). Also the neutron creation and propagation through the shield were studied in detail and compared to experimental data (e.g. \cite{mybyby}, \cite{taupprocessing2015} and \cite{myphd}). 

The geometry of all relevant parts within the cryostats as well as the full shield are implemented in detail in the simulation code. This includes the dimensions, material compositions and densities. The geometry of room A-408 as well as parts of the reactor building including the reactor core were modeled as well for the simulation of the reactor-correlated background contributions as described in \cite{neutronpub}.

The applied physics processes in Geant4 are steered by the so-called physics list. For an accurate description of low energetic electromagnetic interactions as relevant here the "Livermore" physics modules are used, this includes the EADL database for X-ray emissions. Neutron propagation below 20\,MeV, which covers all neutrons simulated within this publication, are calculated with the "Neutron High Precision Models (NeutronHP)". The secondary production cuts, that describe the limit below which an energy deposition is regarded as a single energy deposition and the creation of secondary particles stops, were lowered from the standard values to 1.2\,keV$_{ee}$ for $\gamma$-rays and 850\,eV$_{ee}$ for electrons and positrons within Ge.


In the simulation output, all energy depositions within the four Ge diodes are registered including the coordinates of the interaction sites as well as the particle type. The number of neutrons entering through the surface of the diodes and their energies are also noted. 

\begin{figure} 
    \centering
    \includegraphics[width=0.48\textwidth]{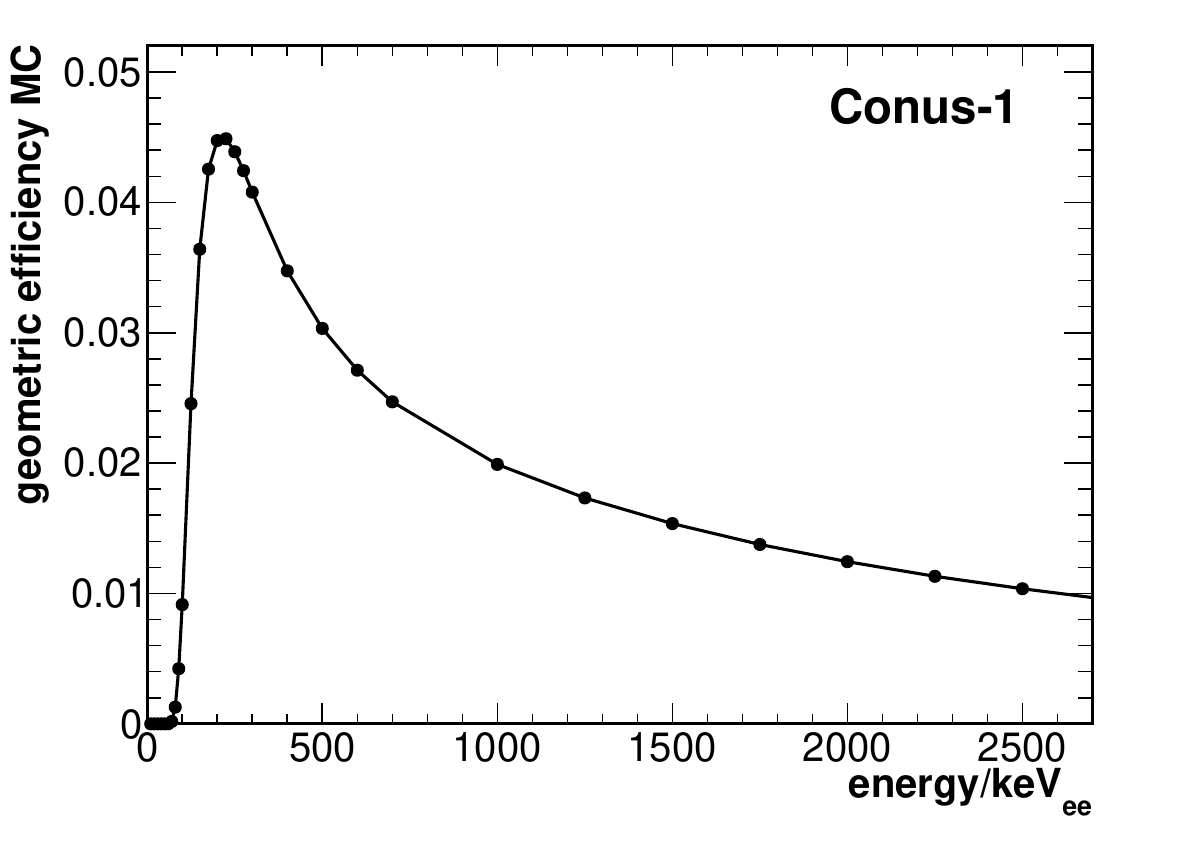}
    \caption{Geometric detection efficiency of the MC simulation for $\gamma$-radiation. To determine the curve, a mono-energetic, isotropic $\gamma$-ray point source was simulated directly on top of the cryostat end cap. It becomes apparent, that all radiation below $\sim$50\,keV$_{ee}$ is blocked by the cryostat, while the efficiency peaks at $\sim$200\,keV$_{ee}$.}
    \label{fig:chapter2_mcefficiency}
\end{figure}

The spectral shape as well as the number of registered events do not only depend on the type and location of the sources, but also on the size and geometry of the detector as well as on the energy collection processes within the diode and in the electronics chain. The latter is included in a post-processing routine, while the geometric detection efficiency is taken into account by the MC simulation. Figure \ref{fig:chapter2_mcefficiency} depicts the ratio of the number of registered $\gamma$-rays in the peak compared to the number of started particles for a mono-energetic isotropic $\gamma$-ray point-source on top of the cryostat end cap (exemplary for C1). The efficiency peaks around 200\,keV$_{ee}$ and decreases towards higher energies due to the increased probability for a $\gamma$-ray to escape the diode before depositing all of its energy. At energies below 50\,keV$_{ee}$ all of the radiation is absorbed by the Cu between the source and the diode. This means that any observed background lines at energies below must arise from within the cryostat end cap.

During the post-processing of simulation data, the following adaptions to convert the output to the read-out Ge signals are included: (1) quenching effect for nuclear recoils, (2) diode surface effects, (3) trigger efficiency and energy resolution and (4) DAQ specific $\mu$-veto dead time.

Quenching (1) means that for nuclear recoils not all energy is deposited as ionization energy, but some is converted to phonons instead. However, Ge spectrometers operated at liquid nitrogen temperatures are only sensitive to ionization energy depositions. The effect has to be included in the MC simulations by hand. The Lindhard theory with adiabatic corrections as described in \cite{scholzquenching} is applied on the individual nuclear recoil energy depositions before adding them up.

\begin{figure*}[h]
\begin{minipage}[h]{8.2cm}
	 \centering\vspace{0cm}
\includegraphics[width=0.99\textwidth]{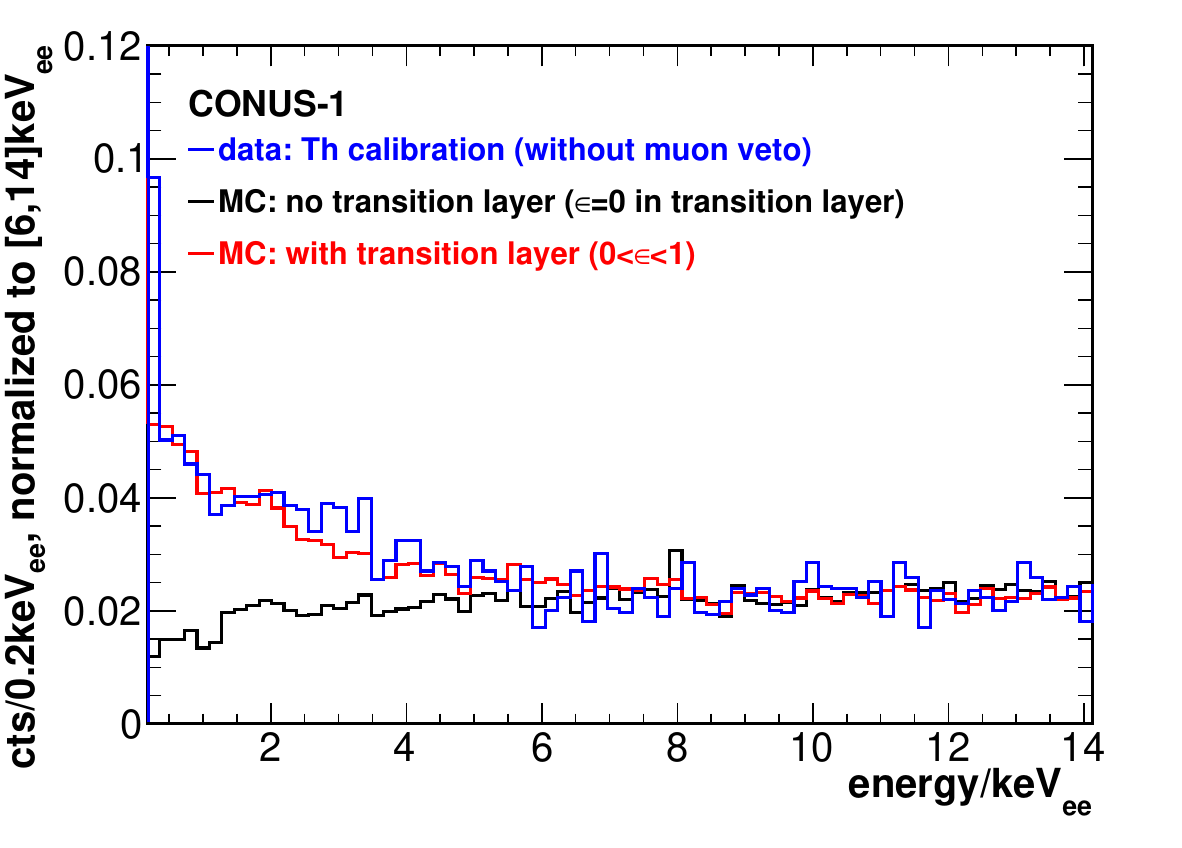}
\caption{Comparison of $^{228}$Th data calibration data of C1 collected at KBR in the lowE regime to MC simulations. The black spectrum corresponds to the simulation without an implemented transition layer, for the red spectrum the charge collection efficiency was modeled in the post-processing of the MC.  }
	\label{chap3:slowpulses_th}
\end{minipage}
\hspace{0.9cm}
\begin{minipage}[h]{8.2cm}
	 \centering\vspace{0cm}
	\includegraphics[width=0.99\textwidth]{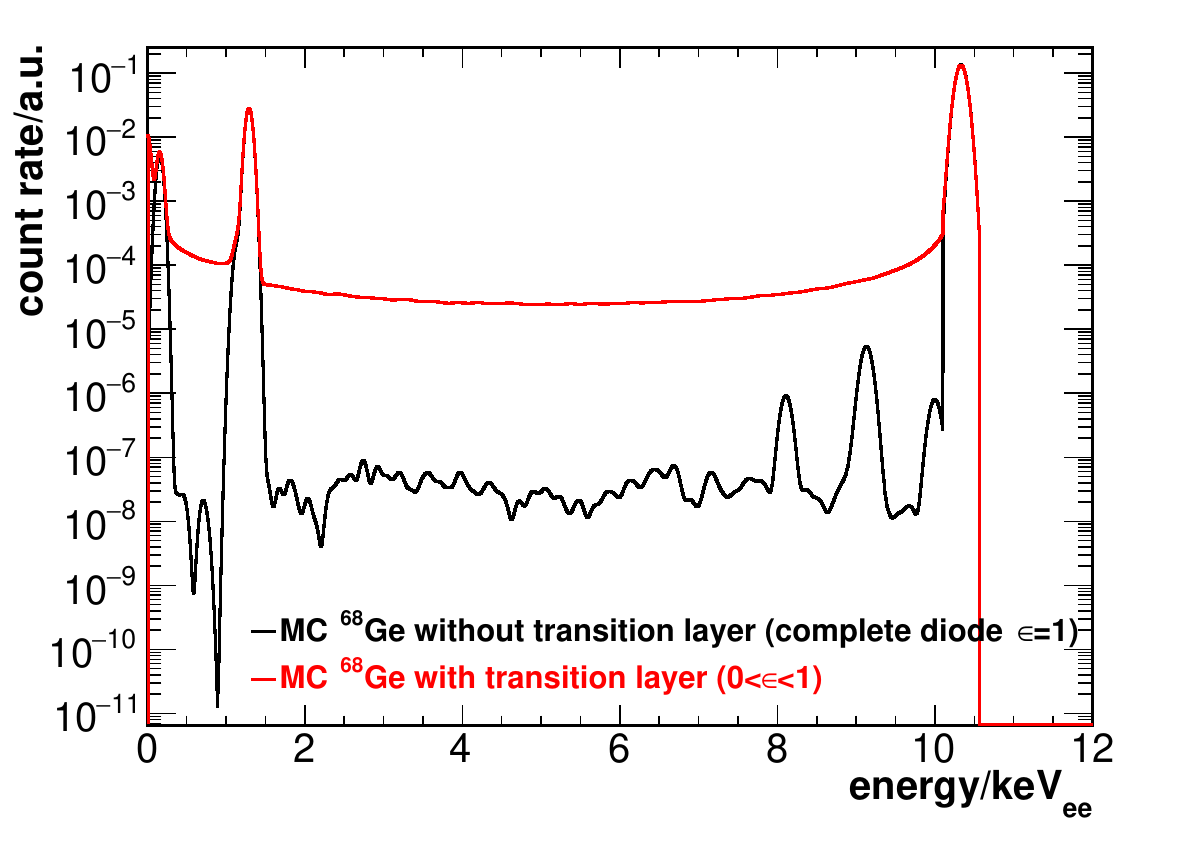}	
	\caption{MC simulation of the decay of $^{68}$Ge within the C1 diode. The continuum between the lines is almost exclusively made up of slow pulses that were created in the transition layer. This is confirmed by the black spectrum, for which neither the transition nor the dead layer was included in the post-processing of the MC and consequently there is nearly no continuum between the peaks.}		
	\label{chap3:slowpulses_ge}
\end{minipage}
 \end{figure*}

Moreover, it has to be taken into account that at the surface of the p-type cylindrical diode (2), there are areas where no or not all energy depositions are registered by the readout electrode (see zoom on the diode displayed in Figure \ref{fig:chapter2_conusshield} (IV)), which has an impact on the normalization and spectral shape of the MC. The passivation layer on the contact side of the diode (thickness $\mathcal{O}(100\,\text{nm})$) as well as the dead layer on all other sides (thickness $\mathcal{O}(1\,\text{mm})$) are fully insensitive (charge collection efficiency $\epsilon=0$). Close to the dead layer within the diode the charge collection efficiency gradually increases ($0<\epsilon<1$) until all deposited energy is collected within the active volume ($\epsilon=1$).
The loss of energy in the transition layer in between results in a shift of the respective signal towards lower energies. These pulses are called slow pulses because the underlying process of recombination of free charges, that results in the energy loss, also delays the collection of the charges arriving at the p+ contact. The active volume as well as the transition layer are modeled by comparing data collected with radioactive sources ($^{241}$Am, $^{228}$Th) to the respective MC simulations as outlined in \cite{detectorpub}. The active volume of all four detectors, evaluated from a comparison of the ratio of various $\gamma$-lines from a $^{241}$Am source to a MC simulation, corresponds to 91-95\% of the Ge crystal volume. 
Approximately $\sim$20\% of the remaining volume makes up the transition layer. The impact of the transition layer can be observed explicitly in the energy range directly below peaks (exemplary see 59.54\,keV peak of $^{241}$Am in Figure 4 of Ref. \cite{detectorpub}) and close to the noise threshold of the detectors. The latter is depicted in Figure \ref{chap3:slowpulses_th} for a $^{228}$Th source calibration measurement, exemplary for C1, in comparison to the MC simulation. For the red curve, a model of the transition layer is included, while for the black curve the charge collection efficiency within the transition layer was set to zero. Each energy deposition within the transition layer including high-energetic ones of several 100\,keV$_{ee}$ is diminished by the charge collection efficiency smaller than one and dropping close to zero in proximity to the dead layer, which leads to the increase towards lower energies. 
A sigmoidal function, comparable to \cite{scholzquenching}, with two parameters is used in the post-processing of the MC simulation to describe the charge collection efficiency within the transition layer. A third parameter is required to crop the efficiency to zero close to the diode surface to get a match between the MC simulation and the experimental data (blue curve).

The impact of the transition layer can be seen as well in Figure \ref{chap3:slowpulses_ge}, where the MC simulation of the decay of $^{68}$Ge within the diode is presented. The continuum of the red spectrum below the lines is nearly fully made up out of slow pulses created at the surface of the diode. The black spectrum depicts the MC simulation with no reduction in the charge collection efficiency included at all.

Another kind of loss occurs within the electronics chain (3). Close to the noise threshold of the detectors, the capability of the DAQ system to register and reconstruct events diminishes towards lower energies. The effect can be quantified by inserting artificially generated signals into the electronics chain and counting the number of signals in the DAQ output. The behavior is depending on the individual detector and Lynx DAQ settings.
For RUN-1, a slight reduction of the efficiency from 100\% at high energies to about 95\% around 300\,eV$_{ee}$ has been observed, while for RUN-2 due to changes in the DAQ settings the efficiency stays at 100\% down to below 250\,eV$_{ee}$ \cite{detectorpub}. Thus the impact on the spectral shape is minor, but nevertheless a correction was applied.   
The detector response also takes into account the variations in the detected energy due to statistical fluctuations in the creation of electron-hole pairs as well as noise induced by the electronics chain. This leads to a smearing of the energy of each event. It is included by folding the MC spectrum with a Gaussian function, whose width corresponds to the energy-dependent energy resolution of the individual detectors. Only a minor impact is expected on the continuum outside of the peaks.

Finally, the $\mu$-veto induced dead time (4) depending on the reactor status as stated in Section \ref{chapter2_shield} is applied as scaling factor to the MC model.

\section{Background decomposition}
\label{chapter4}

In the following section, the different background contributions are described. The origin of the contribution, the potential reduction by the shield, variations over time, and the impact on the overall background count rate are discussed.

\subsection{Muon-induced background inside the CONUS shield}
\label{subsec:muons}
\begin{figure}[h] 
    \centering
    \includegraphics[width=0.48\textwidth]{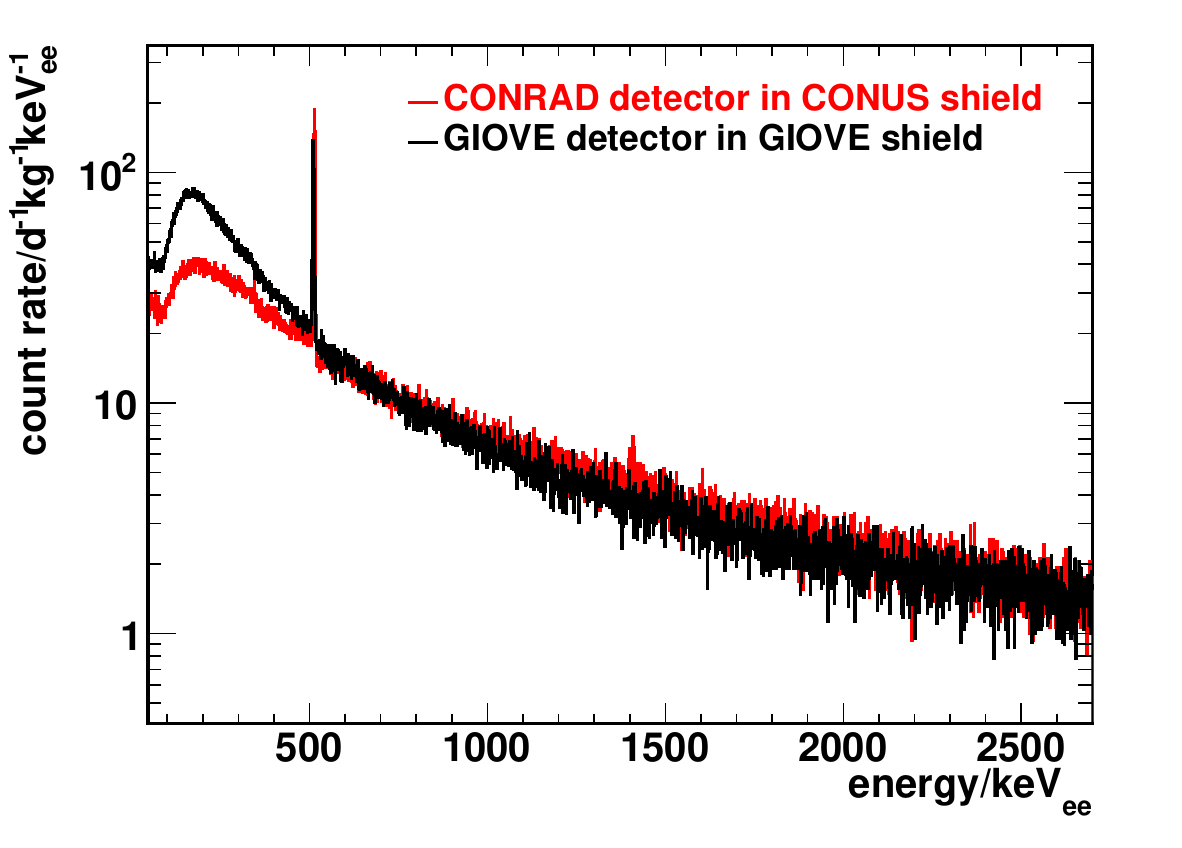}
    \caption{Comparison of $\mu$-induced background in the GIOVE (black) setup to the spectrum detected with the CONRAD detector within the CONUS shield (red). For GIOVE the innermost layer is made out of Cu, while for the CONUS shield the material is Pb. This leads to an improved suppression of the {\it bremsstrahlung} continuum below 500\,keV$_{ee}$. Figure previously published in \cite{taupproceeding2017}.}
    \label{fig:chapter4_muongiovevsconrad}
\end{figure}

Many compact low background experiments need high-density materials such as Pb to shield against $\gamma$-radiation from the environment. Within the CONUS shield 25\,cm Pb in all directions are included. However, this part of the shield is also a target for muons to create secondary radiation consisting of an electromagnetic component as well as neutrons. For experiments located under a shallow overburden ($<$100\,m~w.e.) this background can quickly become dominant over other background sources. In low energy threshold applications, the {\it bremsstrahlung} continuum below 500\,keV$_{ee}$ is shaped by the choice of the shield material close to the diode. For a larger atomic number $Z$ more {\it bremsstrahlung} is created ($\sim Z^{2}$), however the self-shielding increases as well ($\sim Z^{5}$). This results in a lower $\mu$-induced background for an innermost shield layer made out of Pb (Z=82) compared to a Cu (Z=29) layer. This finding was implemented in the CONUS shield and confirmed in the comparison of the background data collected inside the CONUS shield at LLL with CONRAD (innermost layer made out of Pb) to the data from the GIOVE setup located at the same laboratory in a distance of 2\,m (innermost layer made out of Cu \cite{giovepub}) in Figure \ref{fig:chapter4_muongiovevsconrad}. In both cases the diodes have a comparable mass of about 2\,kg.
The background spectra collected with the CONUS detectors is higher by about one third due to the smaller detector mass of about 1\,kg each and the resulting higher Compton continuum.

Data collected without the activation of the $\mu$-veto are completely dominated by the $\mu$-induced component. Such kind of data were collected at LLL and KBR with the CONUS detectors. The spectral shape is identical and the spectra only differ overall by a factor of 1.62 (see Figure 7 in \cite{neutronpub}). In this way, from the known overburden of the LLL of 15\,m~w.e., an effective overburden of 24\,m~w.e. was derived. In both cases, the same processes are responsible for the creation of the muon-induced background.

\subsubsection{MC simulation of muon-induced background inside the CONUS shield}

The MC simulation of the $\mu$-induced background is set up for the LLL and thoroughly tested as described in Ref. \cite{mybyby}. 
Positive and negative muons are started from the walls of the laboratory and propagated through the shield. The spectral and angular distributions are calculated from the ones at Earth's surface \cite{bugaev,reyna}. To reproduce the data collected at KBR, the scaling factor 1.62 from the previous section is used. This results in a muon flux of $\sim$60\,m$^{-2}$s$^{-1}$ used for the normalization.

\begin{figure*}[h]
\begin{minipage}[h]{8.2cm}
	 \centering\vspace{0cm}
\includegraphics[width=0.99\textwidth]{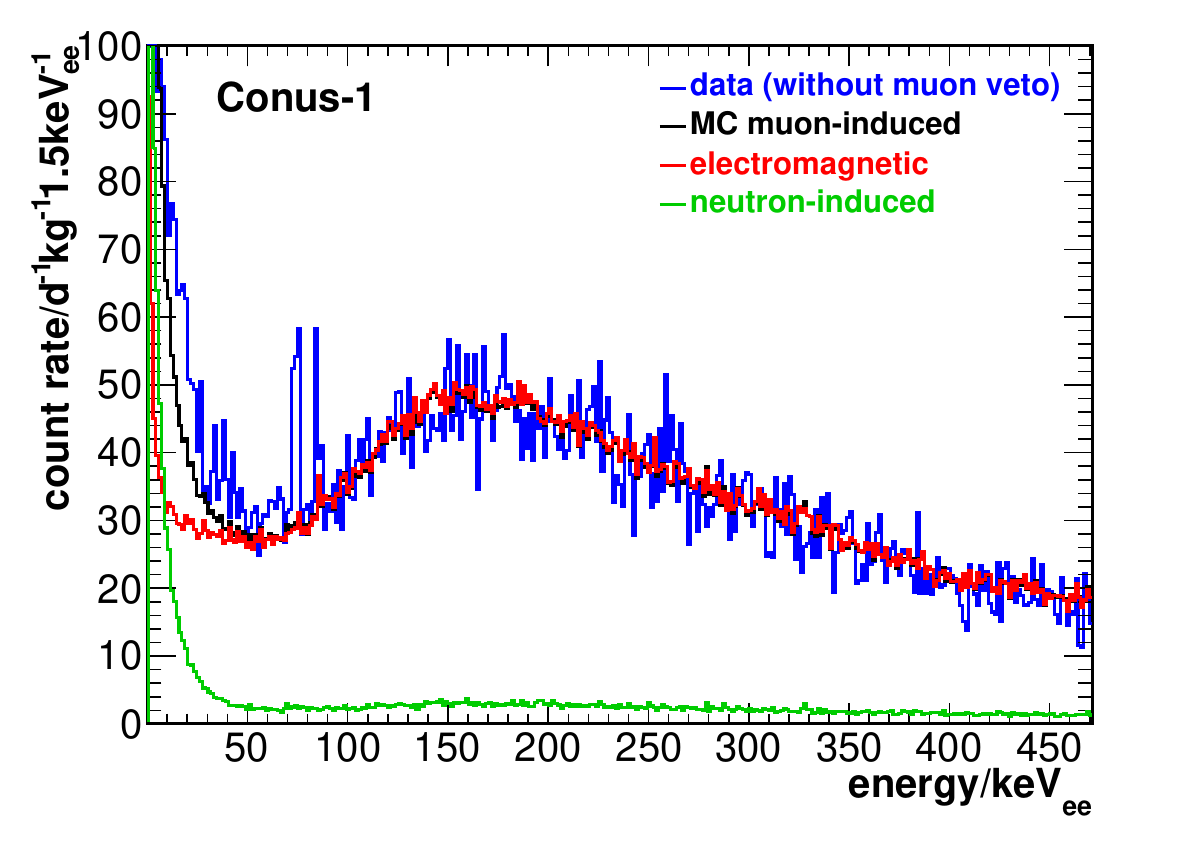}
\caption{Comparison of MC of $\mu$-induced signals to highE data of C1 without applied $\mu$-veto (corrected for other not $\mu$-induced background). Above 100\,keV$_{ee}$ an excellent agreement is achieved. The red spectrum corresponds to the electromagnetic contribution, the green spectrum relates to the neutron-induced background. }
	\label{chap4:highEmuoninducedMC}
\end{minipage}
\hspace{0.9cm}
\begin{minipage}[h]{8.2cm}
	 \centering\vspace{0cm}
	\includegraphics[width=0.99\textwidth]{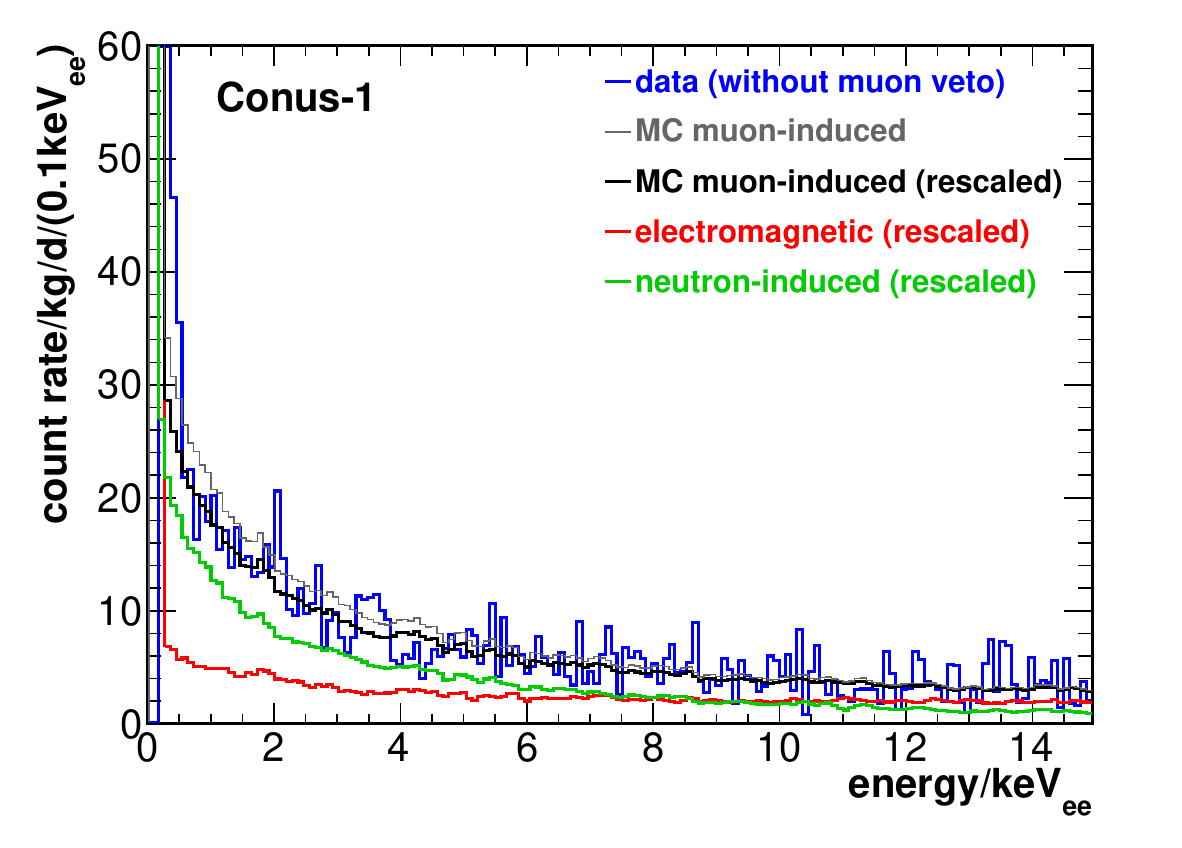}	
	\caption{Comparison of MC of $\mu$-induced signals to lowE data of C1 without applied $\mu$-veto (corrected for other not $\mu$-induced background). A good agreement is observed. The red spectrum corresponds to the electromagnetic contribution, the green spectrum relates to the neutron-induced background. By a minor rescaling, the agreement can be improved. }		
	\label{chap4:lowEmuoninducedMC}
\end{minipage}
 \end{figure*}

The MC simulation is validated directly by a comparison to the data collected without $\mu$-veto (corrected for the not $\mu$-induced background). The outcome is displayed in Figure \ref{chap4:highEmuoninducedMC} and Figure \ref{chap4:lowEmuoninducedMC} as well as in Table \ref{chap4tab:muonmctodatacomparison}. 
Within the MC simulation it is possible to distinguish between all energy depositions depending on whether a neutron is involved or not. 
In the figures for both energy ranges the neutron-induced energy depositions are shown in green, while the electromagnetic contribution is displayed in red. It is evident that the {\it bremsstrahlung} continuum belongs to the electromagnetic part, while the neutrons mostly contribute towards lower energies via nuclear recoils affected by quenching. Close to the ROI below 10\,keV$_{ee}$ the neutron-induced signal correspond to $\sim$80\% of the total $\mu$-induced contribution.

\begin{table}[h]
\begin{center}
\begin{footnotesize}
\begin{tabular}{llll}
\hline
integral                  &  meas. data &MC&MC  \\  
range                &  & & (rescaled) \\  
                         & d$^{-1}$kg$^{-1}$ & d$^{-1}$kg$^{-1}$ & d$^{-1}$kg$^{-1}$\\ \hline
$[0.4,1]$keV$_{ee}$        & 145$\pm$10& 155$\pm$20&130 \\
 $[$2,8$]$keV$_{ee}$     & 460$\pm$20&  500$\pm$50 &440\\
 $[$10,100$]$keV$_{ee}$  & 2300$\pm$30 &1900$\pm$200 &-- \\
  $[$100,440$]$keV$_{ee}$   & 7700$\pm$60& 7800$\pm$800&--\\
 \hline
 
\end{tabular}
\end{footnotesize}
\caption{Comparison of MC simulation of the muon-induced background to data collected without $\mu$-veto for C1. Above 100\,keV, an excellent agreement is achieved. For the background model, the lowE MC was rescaled to improve the agreement (see also Figure \ref{chap4:lowEmuoninducedMC}). All uncertainties on the simulated data are statistical. \label{chap4tab:muonmctodatacomparison}}
\end{center}
\end{table}

Above 100\,keV$_{ee}$ an excellent agreement is achieved, while below minor discrepancies are found. Especially the Pb X-rays at 70-85\,keV are not observable in the MC. It was shown that there is most likely an underproduction of X-rays in Pb in the MC\footnote{In a setup without any Cu parts, it becomes evident that the Pb X-rays are created in the MC. However, the amount of Cu required to suppress the lines is much smaller than the amount used in reality indicating an underproduction in Geant4.}. This, in combination with the well validated neutron fluence rate at the diode described in the next section, points towards the electromagnetic contribution being responsible for the observed small difference. For the background model in the lowE range, the electromagnetic contributions was increased by 10\% and the neutron-induced spectrum was reduced by 20\% to improve the agreement (see black spectrum in Figure \ref{chap4:lowEmuoninducedMC}). This rescaling is purely functional and not used to derive conclusions on the underlying physics.

As shown in Figure \ref{fig:chapter2_bkgsuppression_conusshield} the $\mu$-induced background dominates over all other contributions without the $\mu$-veto applied. The $\mu$-veto strongly diminishes this background contribution. To consider the impact of the $\mu$-veto in the MC, a scaling factor is used under the assumption of an equal performance of the $\mu$-veto over the full spectral range. Indications on the $\mu$-veto efficiency are given by the analysis of the reduction of the 74.97\,keV Pb X-ray in the vetoed spectrum. It is suppressed by $\sim$94\% for C1. The final efficiency in the background model is evaluated from comparing the combination of all scaled MC background contributions to the measured data as discussed in Section \ref{chapter6}. It is estimated to be (97.0$\pm$0.5)\%.

\subsubsection{Muon-induced neutrons inside the CONUS shield}

\begin{figure*}[h]
\begin{minipage}[h]{8.2cm}
	 \centering\vspace{0cm}
\includegraphics[width=0.99\textwidth]{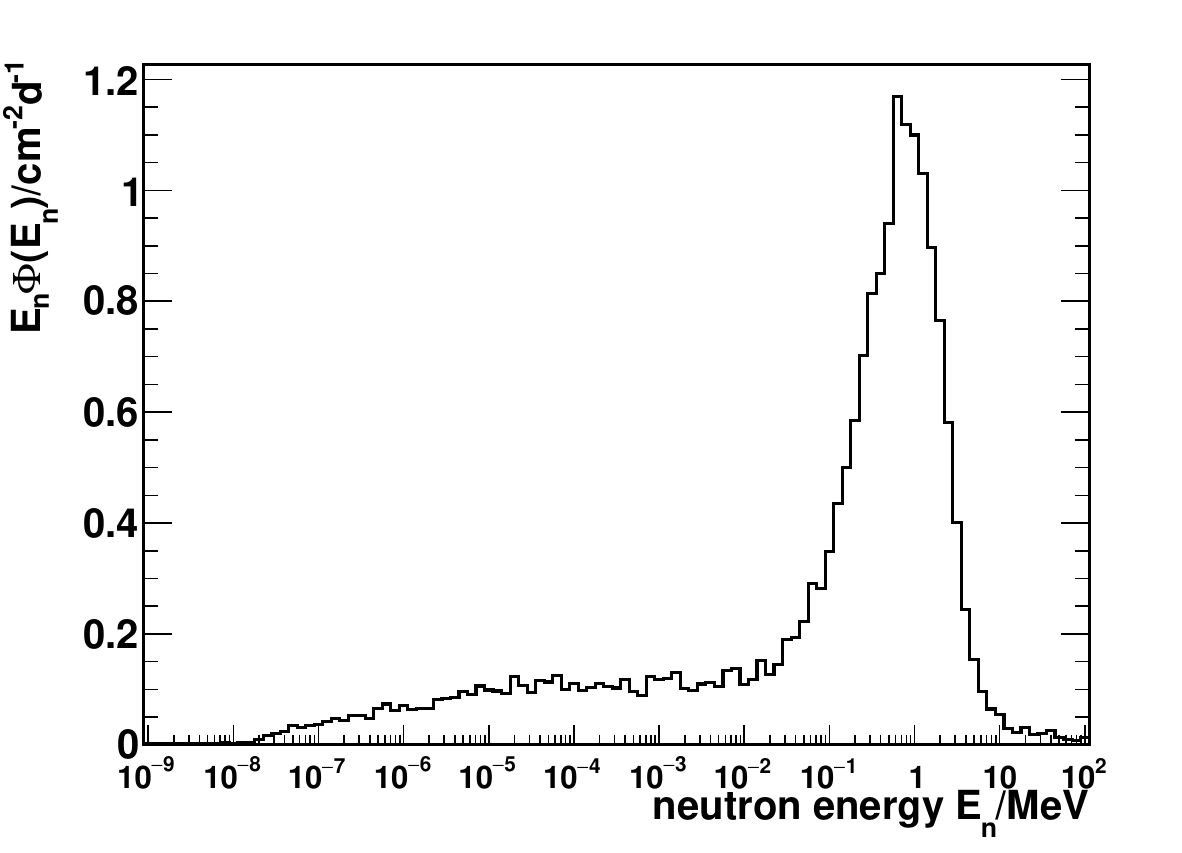}
\caption{Simulated muon-induced neutron fluence rate at a CONUS diode within the CONUS shield at KBR. Most of the neutrons are fast with energies around 1\,MeV that create recoils in Ge.}
	\label{chap4:mc_neutronfluencerateatdiode}
\end{minipage}
\hspace{0.9cm}
\begin{minipage}[h]{8.2cm}
	 \centering\vspace{0cm}
	\includegraphics[width=0.99\textwidth]{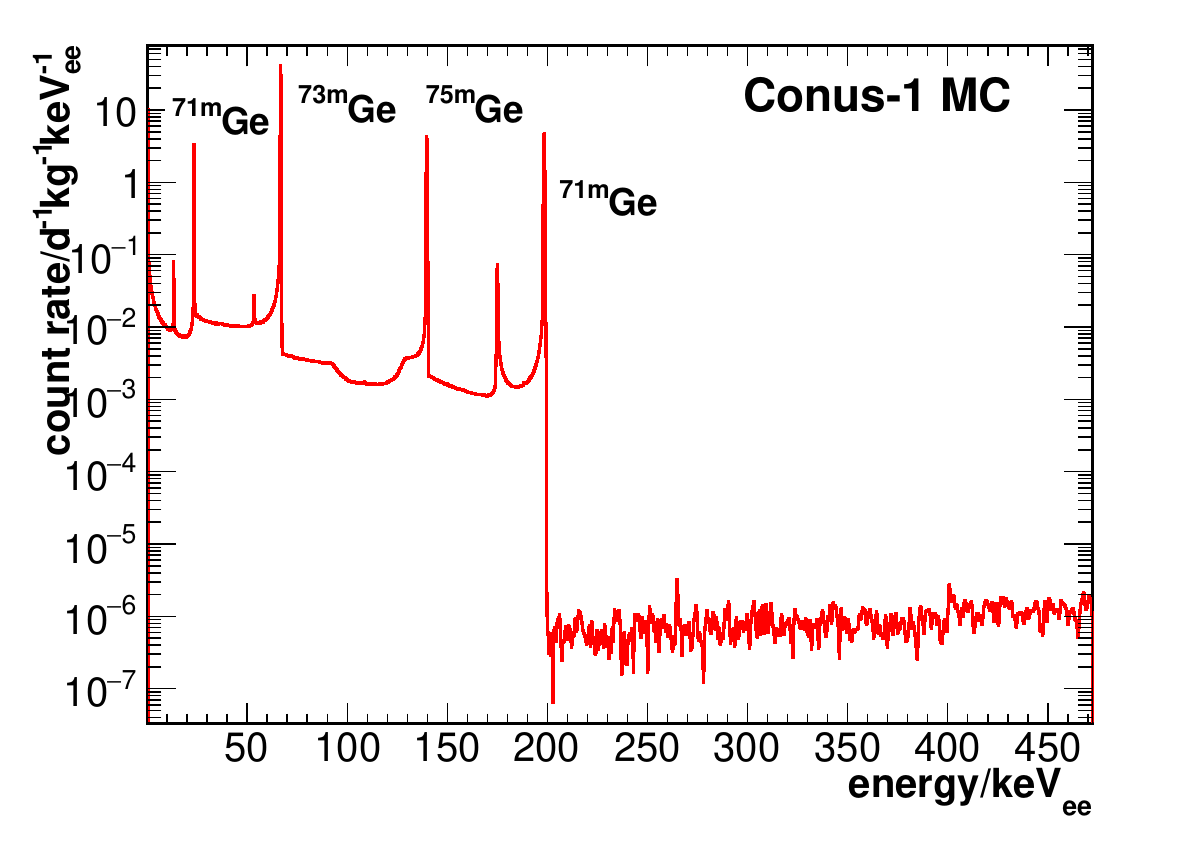}	
	\caption{MC of the decay of the three metastable states in Ge which induce visible background lines in the CONUS data. The isotopes are created by neutron capture of $\mu$-induced neutrons and have a half-life longer than the $\mu$-veto window. The MC spectra are scaled by the measured line count rates during data collection of RUN-1, for each isotope at least one line is observed.}		
	\label{chap4:mc_metastablestates}
\end{minipage}
 \end{figure*} 

The MC predicts a total $\mu$-induced neutron fluence rate of (18$\pm$2)\,cm$^{-2}$d$^{-1}$ passing through the surface of the each CONUS diode at KBR. The validity of the simulation was confirmed by an independent measurement with a Bonner Sphere inside the CONUS shield at LLL \cite{mybyby}. Most of the neutrons are fast with energies around 1\,MeV, only about 6\% of the neutrons are thermal (see Figure \ref{chap4:mc_neutronfluencerateatdiode}). The fast neutrons below 100\,MeV mostly interact with the Ge by elastic and inelastic scattering resulting in nuclear recoils without and with an additional emission of $\gamma$-radiation. The neutron energy is not high enough to induce spallation reactions resulting in long-lived radioisotopes (see Section \ref{subsec:cosmicactivation}). The thermal neutrons dominantly interact via neutron capture. Ge is the most relevant target material within the shield due to the cross sections as well as the geometric detection efficiency. Some of the capture products have a half-life longer than the $\mu$-veto window. Thus, the respective peaks become apparent in the background spectra.  
These are the metastable states $^{71m}$Ge (T$_{1/2}$=20.4\,ms), $^{73m}$Ge (T$_{1/2}$=0.499\,s) and $^{75m}$Ge (T$_{1/2}$=47.7\,s). In Figure \ref{chap4:mc_metastablestates} all of these decays were simulated to show the signatures observed in the detectors\footnote{The deexcitation of $^{73m}$Ge occurs in two steps of 53.4\,keV and 13.3\,keV, the level in between has a half-life of 3\,$\mu$s. This short half-life can lead to a summation of the lines to 66.7\,keV with a tail towards lower energies. The observed shape of the summation structure and the remaining count rate in the 13.3\,keV line depends on the used DAQ. This effect is not included in the post-processing of the MC, as it does not impact the physics analyses in the lowE regime.}. The lines are visible in the measured data, e.g. in the blue spectrum of Figure \ref{fig:chapter2_bkgsuppression_conusshield}. From the MC it becomes evident, that the impact of this background is as low as (0.05$\pm$0.01)\,d$^{-1}$kg$^{-1}$ in [0.4,1]\,keV$_{ee}$ (measured overall background: 5-15\,d$^{-1}$kg$^{-1}$), while being (24$\pm$2)\,d$^{-1}$kg$^{-1}$ in [11,440]\,keV$_{ee}$ (measured overall background: 400-800\,d$^{-1}$kg$^{-1}$ (detector-dependent)).

While the cross section for neutron capture is highest at thermal neutron energies, there is also a finite probability for neutron capture to occur at higher energies. Thus, from the measured line count rates, branching ratios, detection efficiencies, the energy-dependent cross section and the shape of the energy spectrum at the diode, it is possible to indirectly derive (with the method described in more detail in \cite{mybyby}) the neutron fluence rate at the diode. This is an additional possibility to validate the neutron fluence rate at the diode derived from MC. From the reaction $^{74}$Ge(n,$\gamma$)$^{75}$Ge a neutron fluence rate of (20$\pm$3)\,cm$^{-2}$d$^{-1}$ was evaluated in excellent agreement with the MC result.   

Moreover, in the neutron capture processes also the long-lived $^{71}$Ge isotope with a half-life of T$_{1/2}$=11.4\,d is created. It decays via X-ray emission at the same energies as $^{68}$Ge (T$_{1/2}$=271.0\,d), which is activated via spallation processes only possible above ground. From the time development of the count rate of the K-shell X-ray at 10.37\,keV emitted by both isotopes, an {\it in situ} activity of (20$\pm$2)\,d$^{-1}$kg$^{-1}$ for $^{71}$Ge can be derived (see Figure \ref{fig:chapter4_10p4keVlinedecay}). This is discussed in detail in the context of the cosmogenic activation in Section \ref{subsubsec:cosmicdata}.

\subsubsection{Muon-induced neutrons in the concrete of the reactor building}
 
Neutrons are not only induced by muons inside the Pb of the shield, but also within the steel-enforced concrete of the building.
These neutrons were detected together with the reactor-correlated neutron background in the Bonner sphere screening campaign before the setup of the experiment \cite{neutronpub}. The detected spectrum was propagated through the shield in the MC simulation resulting in a neutron fluence rate of (0.126$\pm$0.005)\,cm$^{-2}$d$^{-1}$ at the diodes, which is about a factor of eight below the $\mu$-induced neutron fluence rate inside the detector chamber after application of the $\mu$-veto under the assumption of a veto efficiency of $\sim$97\%. The spectral shape is comparable to the $\mu$-induced neutrons within the shield, i.e. this background contribution also results in neutron-induced recoils within the ROIs.
For the energy range of [0.4,1]\,keV$_{ee}$ (0.4$\pm$0.1)\,d$^{-1}$kg$^{-1}$ are expected from the MC and in [2,8]\,keV$_{ee}$ there are (0.8$\pm$0.1)\,d$^{-1}$kg$^{-1}$, at higher energies the contribution is negligible.

\subsection{Cosmogenic activation}
\label{subsec:cosmicactivation}
Cosmogenic activation of detector and shield materials can occur at Earth's surface or {\it in situ} at the experimental site within the shield. 

At Earth's surface the hadronic component of cosmic rays (mostly protons and neutrons) creates long-lived isotopes in spallation processes. For example, for neutrons with energy above 20\,MeV \cite{Ge68_activation} the following reaction occurs: 
\begin{equation}
^{70}\text{Ge} + n(\text{fast}) \rightarrow ^{68}\text{Ge} + 3n    
\end{equation}
Neutron capture such as $^{70}$Ge(n,$\gamma$)$^{71}$Ge can happen as well. This is the dominant Ge activation process within the CONUS shield, where the neutron fluence rate is much less energetic (see Figure \ref{chap4:mc_neutronfluencerateatdiode}). 

For CONUS, the activation of the Ge diodes and the Cu parts of the cryostats is most relevant due to the high activation cross sections of the materials and the close distance to the detector.
In Table \ref{tab_cosmicactivation_isotopes} the relevant isotopes are listed. They were selected according to their half-lifes, production rates and signatures of the decay radiation. The only possibility to avoid this contamination is to keep the materials underground ($>$10\,m~w.e. ideally) as much as possible and to achieve a low neutron fluence rate within the shield. 

In the lowE range of the CONUS detectors, only lines emitted within the Ge can be observed due to the shielding ability of the cryostat (see Figure \ref{chap4:mc_cosmics_lowE}). These are X-rays following electron-capture (EC) decays of the isotopes with Z$<$40. There are K shell lines (around 10\,keV$_{ee}$) and L shell lines (around 1\,keV$_{ee}$). For the background within the ROI, also $\beta$-decays, slow pulses and Compton continua from higher energetic $\gamma$-lines of other activation products can be relevant.

In Section \ref{subsubsec:cosmicdata}, the visible background lines in the CONUS data are studied and production rates are derived for a comparison to literature. In Section \ref{subsubsec:cosmicmc}, the contribution to the overall background and in the ROIs is estimated with the help of MC simulations for the contribution inducing line background as well as for those without visible lines in the data. In the latter case, the production rates from literature were used for scaling.

\subsubsection{Cosmogenic activation observed in CONUS}
\label{subsubsec:cosmicdata}

The only clearly visible lines in the lowE range of the data collected at the experimental site at KBR come from the decays of $^{68}$Ge and consecutively $^{68}$Ga, $^{71}$Ge and $^{75}$Zn (see Table \ref{tab_energiescosmogenic} for the energies of the K shell and L shell X-rays and Figures \ref{fig_completebkgmodelc1} to \ref{fig_completebkgmodelc3} for the measured energy spectra). While the excellent energy resolution of the CONUS detectors makes it possible to easily resolve the three K shell lines, the overlapping L shell lines are combined to a single broad peak.

Within the highE range additional lines from the decay of $^{57}$Co within Ge and Cu can be seen. The 121.1\,keV (branching ratio (br)=0.855) and the 136.5\,keV line (br=0.107) are visible, as well as a summation line at about 143\,keV. It is created in the decay occuring within Ge, where the X-rays at 6-7\,keV as well as 14.4\,keV add to the other emission lines. Due to these decay properties, it can be assumed that the contribution to the 121.1\,keV line nearly fully originates from the Cu, while the 143\,keV line is created by decays inside the Ge. For the highE range, moreover the decay of $^{68}$Ga via a $\beta$-decay is relevant, while for $^{65}$Zn the dominant line at 1115.5\,keV (br=0.502) is outside the spectral range of the DAQ.    

For all other isotopes, there are either no background lines within the highE range or the activity is too low for them to become apparent above the continuous background. The overall small cosmogenic-induced contribution is achieved by successful minimization of the total exposure to the hadronic component of the cosmic rays for the shield and detector materials during manufacturing (see Table \ref{tab_cosmicactivation_exposure}). 

\begin{figure}[htb] 
    \centering
    \includegraphics[width=0.45\textwidth]{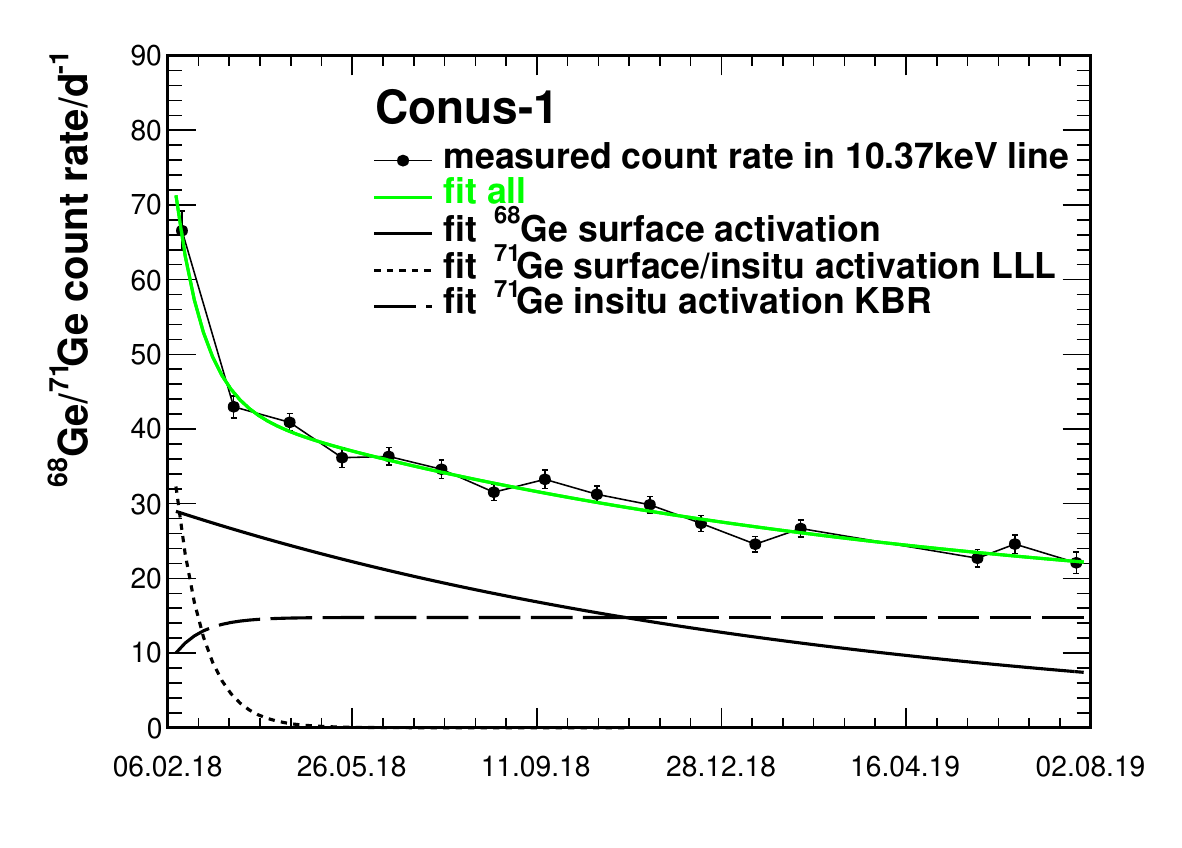}
    \caption{The decay of the count rate of the 10.37\,keV line is observed in the data in monthly bins, exemplary for C1. There are three contributions to the count rate: the decay of $^{68}$Ge (T$_{1/2}$=271.0\,d) induced at the Earth's surface, the decay of $^{71}$Ge (T$_{1/2}$=11.4\,d) induced at the Earth's surface and at LLL and the {\it in situ} activation of $^{71}$Ge within the CONUS shield. Data are fitted successfully with the respective model.\label{fig:chapter4_10p4keVlinedecay}}
\end{figure}

For the study of the decays in all four detectors, an extended data set is used including all data collected during the commissioning phase at LLL, at KBR before the physics data collection, and during RUN-1 and RUN-2. This amounts to a total exposure of approximately 1.5\,kg$\cdot$yr in the lowE range. For the highE range only C1 is available due to the limited number of DAQ channels. All of the isotopes corresponding to the observed background lines have a half-life of the order of one year. Data are split into bins of one month. For each time interval, the count rate within the lines is determined. The data points are depicted exemplary for the 10.37\,keV line of the decay of $^{68}$Ge/$^{71}$Ge for C1 in Figure \ref{fig:chapter4_10p4keVlinedecay}.
 
The development of the count rates over time $t$ is fitted with an exponential function $N_{0}\times$exp(-$\lambda$t) with $\lambda$=ln(2)/T$_{1/2}$. The only free parameter corresponds to the reference count rate $N_{0}$ when the detectors were placed inside the CONUS shield at KBR on January 24, 2018. The half-lifes were fixed to the literature values from Table \ref{tab_cosmicactivation_isotopes}.

\begin{table*}[h]
\caption{Production rates at sea level of all cosmogenic-induced isotopes in Ge and Cu relevant for CONUS. For the literature values mean values and standard deviations are stated (see \cite{cosmicact1} - \cite{cosmicact12} for Ge and \cite{cebriancosmicactivation} for Cu) except for $^{3}$H. For $^{3}$H, the mean value was calculated from the rates determined in \cite{edelweiss2017}, \cite{cosmicact11} and \cite{cosmicact12}, but the uncertainty evaluated by the EDELWEISS experiment is used \cite{edelweiss2017}. From the lines visible in the CONUS background data production rates are derived. The other cases are marked by n.d. (no data). The last column summarizes the expected decay signatures with a significant branching ratio. The applicable radiation is marked with x, and (x) refers to a line background above the highE range. 
\label{tab_cosmicactivation_isotopes}}
\begin{footnotesize}
\begin{tabular}{ll|ll|ll|lll}

\hline\noalign{\smallskip}
 
isotope  & half-life    & \multicolumn{2}{l|}{production rate Ge} & \multicolumn{2}{l|}{production rate Cu} & X-rays & $\gamma$-rays &$\beta$-decays\\
  &   [d]  & \multicolumn{2}{l|}{[atoms~d$^{-1}$kg$^{-1}$]} & \multicolumn{2}{l|}{[atoms~d$^{-1}$kg$^{-1}$]} &   &   & \\
         &                & literature & CONUS data & literature & CONUS data &         &               &             \\ \hline
 $^{68}$Ge & 271.0        & 59$\pm$44  &     [40,70] (III)   &    no        &        no    &   x     &                &             \\       
           &              &              &    200$\pm$30 (I)    &    no         &      no      &       &               &            \\  
 $^{68}$Ga & 0.05         & from $^{68}$Ge &          &  no          &    no        &    x    &  (x)             &  x        \\   
 $^{65}$Zn & 244.0        & 50$\pm$22   &   60$\pm$10         &   no         &     no       &    x      &       (x)        &     x        \\   
 $^{57}$Co &  271.8       & 7.6$\pm$3.9 &   9.0$\pm$1.0         &    55$\pm$19        &  125$\pm$27          &     x   &          x     &         x   \\   
 $^{60}$Co & 1923.6       & 3.9$\pm$1.5 &     n.d.       & 46$\pm$26           &  n.d.       &   x     &           (x)    &         x   \\   
 $^{54}$Mn & 312.2        & 2.6$\pm$1.7 &    n.d.        & 16$\pm$7           &   n.d.         &    x    &       (x)        &           x  \\     
 $^{3}$H   & 4493.9       & 77$\pm$21   &     n.d.      &    no        &   no         &       &               &       x      \\   
 $^{55}$Fe & 1002.7       & 5.8$\pm$2.6 &    n.d.       &     no       &   no         &  x      &               &             \\   
\noalign{\smallskip}\hline
\end{tabular}
\end{footnotesize}
\end{table*}

Only for the 10.37\,keV line the time development is more complex due to the three sources contributing to the count rate: the decay of $^{68}$Ge activated by the hadronic component of the cosmic rays at Earth's surface, the decay of $^{71}$Ge activated before deployment at KBR at locations with no or a smaller overburden and the {\it in situ} activation of $^{71}$Ge inside the CONUS shield at KBR.
The different fit contributions are depicted in Figure \ref{fig:chapter4_10p4keVlinedecay} for C1.

To determine activities from the reference count rates, the geometric detection efficiency of the single isotopes for the Ge and Cu volumes as well as the branching ratios relevant for the decays are required. Both can be determined from the MC simulations. The validation of the branching ratios as implemented in the simulation is described in the next section. The resulting activities per mass of material are stated in Table \ref{tab:cosmogeniclinebackgroundactivity}. 
The Ge diodes have a mass of 1\,kg each. For the closest and most relevant Cu parts with a mass of about 2\,kg, the detection efficiency is diminished by about one order of magnitude compared to the decays within the Ge. 
As a consistency check of the data analysis, the activities of $^{68}$Ge and its decay product $^{68}$Ga can be compared. The activities, which are supposed to be the same, are found to mostly agree within a 2\,sigma uncertainty.

Taking into account the alternating activation and decay of the isotopes starting with crystal pulling until the deployment of the detectors at KBR (for the total exposure of the materials over ground see Table \ref{tab_cosmicactivation_exposure}), it is possible to evaluate production rates of $^{75}$Zn, $^{57}$Co and under additional assumptions $^{68}$Ge from the CONUS data. In Table \ref{tab_cosmicactivation_exposure}, the total exposure of the materials is summarized. While the exposure of the Ge diodes to cosmic radiation was accurately tracked, for the Cu parts it is less well known. However, in both cases the main contribution is attributed to diode production, detector cryostat assembly and electronics optimisation. 
The results of the determination of the production rates are given in Table \ref{tab_cosmicactivation_isotopes} in comparison to the mean literature values. The average value of all four detectors is stated in the table. For most isotopes, large fluctuations in the reported literature values derived from experiments or simulations are found. Therefore, mean values and the respective standard deviations are reported in the table.  
For $^{65}$Zn and $^{57}$Co in Ge, the production rates derived from CONUS data are in excellent agreement with the literature values.

For the mentioned isotopes as for most activation products in the Ge diode, the activation starts after the crystal pulling, as during the crystallization process all impurities are removed from the material \cite{crystalgrowth}. Only $^{68}$Ge and $^{71}$Ge are an exception, since they remain inside the crystal melt.
We are confronted with three different possible scenarios for $^{68}$Ge with its longer half-life: zero-activation at crystal pulling (I) and the more likely partial (II) as well as total (III) activation.
For the two extreme cases (I) and (III) production rates from the CONUS data are derived. For (I) we found (200$\pm$30)\,atoms~d$^{-1}$kg$^{-1}$. In the case (III) a range of [40,70]\,atoms~d$^{-1}$kg$^{-1}$ was derived. In this case, due to the uncertainty of the actual saturation of the crystal, discrepancies between the detectors were found, which are covered by the given interval. The latter one is in good agreement with the mean literature value, indicating that the crystals were activated close to saturation prior to crystal pulling.

For the activation of $^{57}$Co in Cu, a twice as large activation rate than expected is derived from the CONUS data. The discrepancy might at least partially occur from the not fully tracked exposure of the Cu pieces to cosmic radiation. 

While for the physics data collection at KBR only isotopes with a half-life longer than 100\,d are considered as relevant, it is possible to look for isotopes with a shorter half-life and a larger production rate in the data collected two months before the deployment of the detectors at KBR and immediately after the deployment during the commissioning phase. Here, additionally the 11.1\,keV line of $^{73}$As (T$_{1/2}$=80.3\,d) from the activation of Ge was observed resulting in an activity of (1.1$\pm$0.4)\,d$^{-1}$kg$^{-1}$ on January 24, 2018. X-ray line emissions at $\sim$6.4\,keV by $^{56-60}$Co could not be identified above the continuum.

\begin{table*}[btp]
\caption{Comparison of energies and branching ratios of the K and L shell lines of the isotopes clearly visible in the lowE range of the CONUS data.
The energy values from literature, that are applied for the calibration of the CONUS data, are taken from \cite{Bearden-Burr:1967} with uncertainties \cite{Fuggle-Martensson:1980}, while the K to L shell emission ratios are derived from \cite{nucleideorg}. The energy values evaluated from the CONUS data (exemplary for C1 RUN-1) are given as reference.  
The MC description of the decays is based on the EADL database \cite{eadl}. While there is an excellent agreement between MC and literature on the ratio of emission from the different shells, discrepancies for the energy values are found. The two K shell energies for $^{65}$Zn are discussed in detail in the text. \label{tab_energiescosmogenic}}
\begin{tabular}{llllll}
\hline\noalign{\smallskip}
                       &   \multicolumn{3}{c}{energy/keV}           &  \multicolumn{2}{c}{branching ratio K/L}  \\
isotope                &   literature            &MC         &   C1 RUN-1&literature              &    MC\\\hline 
 K shell               &                         &          &         &                        &         \\  
 $^{68}$Ge              &    10.3665$\pm$0.0005  &    10.33  &10.366$\pm$0.001         &    0.133$\pm$0.003      &0.132  \\
 $^{71}$Ge             &    \texttt{"} & \texttt{"}  & \texttt{"}     &    0.119$\pm$0.003      &0.119  \\
 $^{68}$Ga             &     9.6586$\pm$0.0006   &     9.69  & 9.663$\pm$0.005       &    0.111$\pm$0.001      &0.110  \\
 $^{65}$Zn             &     8.9789$\pm$0.0004   &     8.96  & n.d.                   &    0.114$\pm$0.001      &0.110  \\
 
                       &     8.9054              &   n.d.        &  8.933$\pm$0.004       &      n.d. &  n.d.\\    \hline  
L shell                &                         &           &         &                         &        \\
 $^{68}$Ge             &    1.2977$\pm$0.0011    &     1.291 &    1.299$\pm$0.010    &                         &  \\  
 $^{71}$Ge             &    \texttt{"}    &   \texttt{"} &         &                         &  \\
 $^{68}$Ga                      &    1.1936$\pm$0.0004    &     1.252 &        &                         &  \\
 $^{75}$Zn             &    1.0966$\pm$0.0004    &     1.101 &        &                         &  \\
 \noalign{\smallskip}\hline
\end{tabular}
\end{table*}
\subsubsection{MC of cosmogenic induced isotopes}
\label{subsubsec:cosmicmc}
To quantify the contribution to the ROI and especially the decay over time, all relevant radioisotopes from Table \ref{tab_cosmicactivation_isotopes} were simulated. 

For the four isotopes with the observable background lines ($^{68}$Ge, $^{71}$Ge, $^{68}$Ga and $^{65}$Zn) the energy as well as the K to L shell emission branching ratios were validated by a comparison to literature in Table \ref{tab_energiescosmogenic}. For some cases differences up to a few tens of eV are observed, especially for the K shell line of $^{68}$Ge/$^{71}$Ge as well as the L shell line of $^{68}$Ga. However, the discrepancies are considered too small to forego a correction, especially as the lines are not included into the likelihood analyses.  
For the K shell line of $^{65}$Zn the literature of the last decades only provides one energy value of 8.98\,keV. Just in the most updated X-ray analysis \cite{nucleideorg} two energies are quoted, one at 8.98\,keV with known branching ratio and one at 8.90\,keV without known branching ratio. In the MC, only a line at 8.96\,keV is included as stated in the EADL database, which is based on theoretical approaches \cite{eadl}. 
In the past, many Ge-based experiments (e.g. \cite{edelweiss2016}, \cite{malbek}, \cite{cogent}), that were able to observe the lowE range, used exclusively the 8.98\,keV value. On the contrary, the CONUS data prefers a value at 8.93\,keV, which suggests a contribution from both literature values. Since the energy calibration in CONUS relies on a comparison of fitted versus literature values, we applied a larger uncertainty on the literature values of $^{65}$Zn to overcome this discrepancy \cite{detectorpub}. In the background data of the CDEX experiment \cite{cdex} there is also an indication for a preference of a slightly lower energy value of this line. For the ratio of the K and L shell branching ratios an excellent agreement was found in all cases, especially the differences in the decay of $^{68}$Ge and $^{71}$Ge (same emitted energies) are reflected correctly. An even more detailed validation of the decay of $^{71}$Ge is described in \cite{supercdmsGe71}.

The normalization of the MC simulations for these four isotopes was calculated from the activities in Table \ref{tab:cosmogeniclinebackgroundactivity} and the respective models for the time development of the decays. For all other isotopes, the activity derived from the mean values from literature in Table \ref{tab_cosmicactivation_isotopes} and the respective half-life were used instead for normalization. The respective standard deviations dominate the overall uncertainty of this component. 

\begin{figure*}[htb]
\begin{minipage}[htb]{8.cm}
	 \centering\vspace{0cm}
\includegraphics[width=0.95\textwidth]{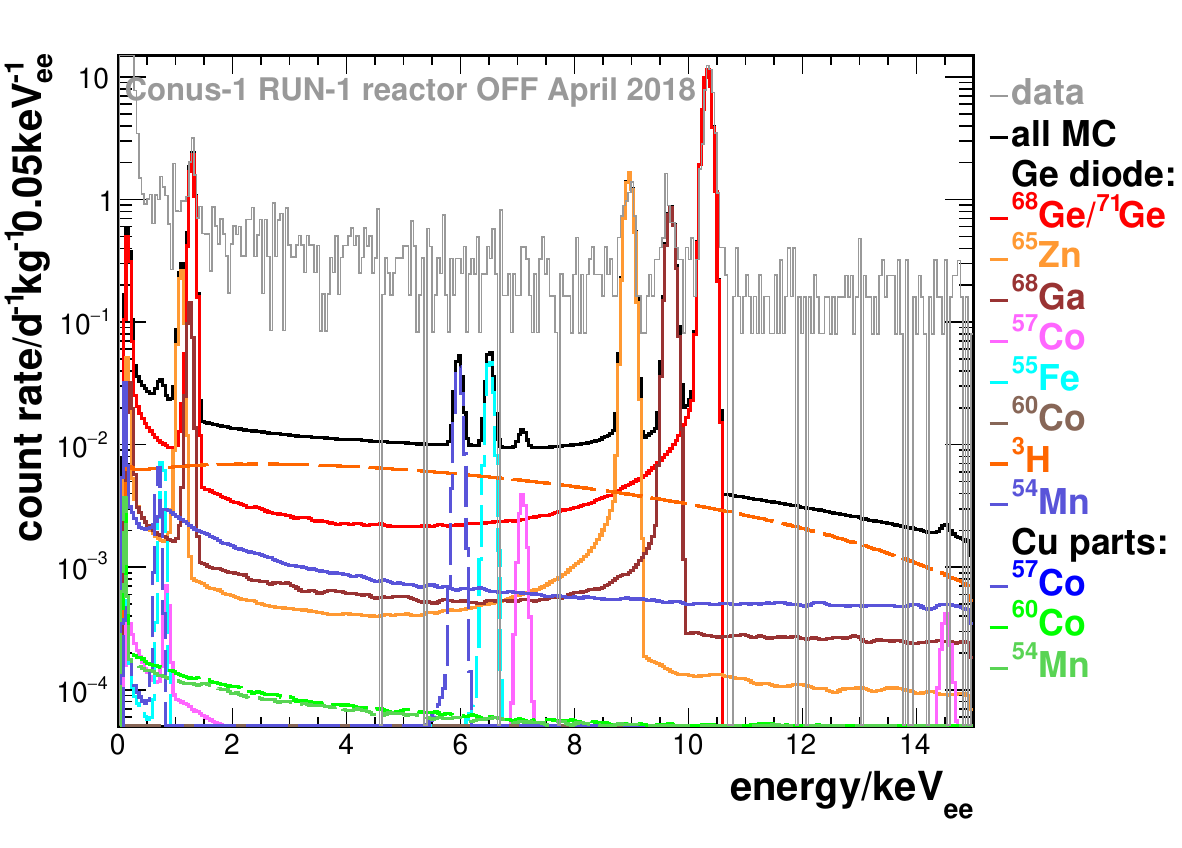}
\caption{Contribution of the decay of cosmogenic-induced isotopes evaluated with the MC simulation in comparison to background data of C1 in April 2018 in the lowE range.  }
	\label{chap4:mc_cosmics_lowE}
\end{minipage}
\hspace{0.5cm}
\begin{minipage}[htb]{8.cm}
	 \centering\vspace{0cm}
	\includegraphics[width=0.95\textwidth]{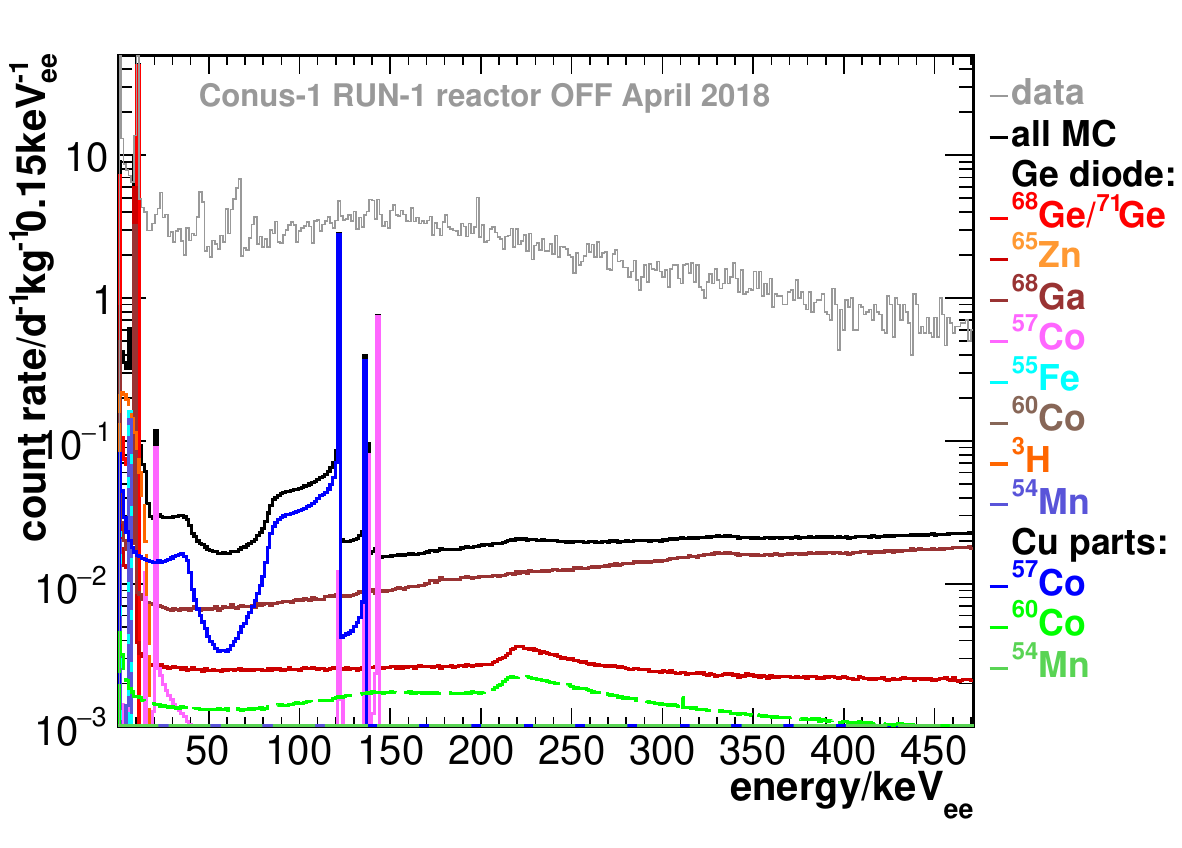}	
	\caption{Contribution of the decay of cosmogenic-induced isotopes evaluated with the MC simulation in comparison to background data of C1 in April 2018 in the highE range. }		
	\label{chap4:mc_cosmics_highE}
\end{minipage}
 \end{figure*}

In Figure \ref{chap4:mc_cosmics_lowE} and Figure \ref{chap4:mc_cosmics_highE} the resulting spectra are depicted in comparison to the background data collected during the outage of RUN-1 in April 2018 for C1. It was found that below the K shell emission lines at 8\,keV$_{ee}$ (within the energy range of various BSM analyses \cite{bsmpaper}) the most relevant contribution is provided by the $\beta$-decay of $^{3}$H. This is followed by the contributions of $^{68}$Ge/$^{71}$Ge, which are also dominant below the L shell lines within the ROI of CE$\nu$NS. 
For these decays the continuum is nearly exclusively populated by slow pulses (see Section \ref{chap3:slowpulses_ge}). Within the highE range, the largest contribution to the continuum is made up of $^{57}$Co decays within the Cu parts close to the diode, followed by the $\beta$-decay emission of $^{68}$Ga. 

In Table \ref{tab:cosmogeniclinebkgcontribution} the total contribution of the \linebreak[4]cosmogenic-induced isotopes is listed in various energy ranges and over time, as predicted by the MC simulation. First of all, it becomes evident that outside of the lines this background contribution is sub-dominant compared to the overall background. Nevertheless, as the CONUS experiment is looking for rare signals revealing themselves in a comparison between reactor OFF and ON data sets, the potential decaying contribution still needs to be evaluated carefully. It turns out that within [0.3,1]\,keV$_{ee}$ a reduction of 0.1-0.2\,d$^{-1}$kg$^{-1}$ is expected between the outages of RUN-1 and RUN-2 more than one year apart (measured overall background 10-15\,d$^{-1}$kg$^{-1}$), while for [2,8]\,keV$_{ee}$ it is a reduction of 0.5\,d$^{-1}$kg$^{-1}$ at most (measured overall background 25-40\,d$^{-1}$kg$^{-1}$). The decrease in the decay rate of the cosmogenic isotopes during the run was considered negligible for the physics analyses of both runs.

\begin{table}[btp]
\caption{Total exposure above ground of the Ge and Cu employed in the CONUS experiment. While for Ge the exposure was tracked exactly, there are larger uncertainties for the Cu parts. During the manufacturing process, all of these materials were kept underground as much as possible to avoid cosmic activation.\label{tab_cosmicactivation_exposure}}
\begin{tabular}{lll}
\hline\noalign{\smallskip}
detector & \multicolumn{2}{l}{exposure above ground [d]} \\
         & Ge    &Cu\\\hline
C1       & 135  &$\sim$116\\
C2       & 99   & $\sim$73 \\
C3       & 66   & $\sim$101\\
C4       & 91   & $\sim$116 \\
\noalign{\smallskip}\hline
\end{tabular}
\end{table}

\begin{table*}[hbtp]
\begin{center}
\begin{footnotesize}
\begin{tabular}{lccccccc}
\hline 

   &\multicolumn{7}{c}{ activity [d$^{-1}$kg$^{-1}$]}     \\\hline
        isotope  & $^{71}$Ge    & $^{71}$Ge (*)  &$^{68}$Ge &$^{68}$Ga &$^{65}$Zn   & $^{57}$Co& $^{57}$Co  \\  
          material. &            Ge             &  Ge                   &   Ge            &       Ge         &    Ge                        &  Ge       &  Cu     \\ 
  mass   &             1\,kg             &        1\,kg             &    1\,kg     & 1\,kg &   1\,kg             &  1\,kg  &  2\,kg    \\ 
location               &    LLL/surface   &     KBR & surface & surface &  surface&surface &  surface \\\hline        
C1         &     128.0$\pm$14.3        &  18.8$\pm$1.3         & 39.2$\pm$2.3    &  46.5$\pm$3.2      &   11.9$\pm$0.5               &    2.5$\pm$0.4  & 17.4$\pm$3.9 \\
C2         &     193.1$\pm$55.0        &  17.6$\pm$1.1         &  24.6$\pm$2.1    &  26.9$\pm$2.7     &    7.8$\pm$0.4                &    n.d.     &   n.d.   \\
C3         &       124.0$\pm$16.3      & 21.0$\pm$1.1          &  20.4$\pm$2.0   &  32.0$\pm$2.4      &    7.9$\pm$0.4                &    n.d.      &  n.d.    \\
C4         &     236.0$\pm$15.2        & 19.4$\pm$1.2          & 21.1$\pm$2.1    &  29.4$\pm$2.8      &   6.6$\pm$0.4                &   n.d.      &   n.d. \\\hline
\end{tabular}
\end{footnotesize}
\caption{Activities in d$^{-1}$kg$^{-1}$ of the cosmogenic-induced isotopes at reference time (deployment at KBR: January 24, 2018) derived from the observed time development of the visible background lines in the CONUS data at KBR, the respective decay models, the branching ratios and MC detection efficiencies. For $^{71}$Ge (marked with (*)) also the constant {\it in situ} activity within the CONUS shield at KBR is given, that will be observed after the decay of the other contributions. The uncertainties are determined by the fit uncertainties, which take into account the statistic uncertainties on the determination of the line count rates. \label{tab:cosmogeniclinebackgroundactivity}}
\end{center}
\end{table*}

\begin{table}[hbt]
\begin{center}
\begin{footnotesize}
\begin{tabular}{llll}
\hline 

    detector \&         &[0.3,1]\,keV$_{ee}$  & [2,8]\,keV$_{ee}$  & [11,440]\,keV$_{ee}$    \\ 
    timestamp         &   d$^{-1}$kg$^{-1}$   &   d$^{-1}$kg$^{-1}$    &   d$^{-1}$kg$^{-1}$           \\\hline
            \multicolumn{4}{l}{C1}         \\   
            Jan. 24, 2018       &1.0$\pm$0.1  &2.7$\pm$0.6 & 14.5$\pm$1.5 \\
            RUN-1 OFF       & 0.5$\pm$0.1 & 1.8$\pm$0.6 & 11.9$\pm$1.3 \\
            RUN-2 OFF        & 0.3$\pm$0.1  & 1.3$\pm$0.6 & 4.4$\pm$0.5 \\
               \multicolumn{4}{l}{C2}         \\   
            Jan. 24, 2018       & 0.9$\pm$0.2  &2.3$\pm$0.6 & 7.3$\pm$1.0 \\
            RUN-1 OFF       & 0.3$\pm$0.1  &1.2$\pm$0.5 & 6.0$\pm$0.8 \\
            RUN-2 OFF         &0.2$\pm$0.1  &0.9$\pm$0.4 & 2.3$\pm$0.3 \\ 
  \multicolumn{4}{l}{C3}         \\   
             Jan. 24, 2018       & 0.6$\pm$0.1  &1.6$\pm$0.3 & 8.6$\pm$1.2 \\
             RUN-1 OFF     & 0.3$\pm$0.1  &1.0$\pm$0.3 & 7.1$\pm$1.0 \\
             RUN-2 OFF       & 0.2$\pm$0.1  &0.7$\pm$0.3 & 2.7$\pm$0.4 \\  
 \hline
\end{tabular}
\end{footnotesize}
\caption{Rate contribution of the decay of the cosmogenic isotopes within various energy ranges during different time periods, as evaluated from the MC. \label{tab:cosmogeniclinebkgcontribution}}
\end{center}
\end{table}

\subsection{$^{210}$Pb within shield and detector cryostats}
\label{subsec:pb}

The CONUS shield contains 25\,cm of Pb to shield natural radioactivity and other external radiation. Specifically, the innermost layer is made out of Pb to profit from the enhanced suppression of the $\mu$-induced \linebreak[4]{\it bremsstrahlung} continuum as detailed in Section \ref{subsec:muons}.

However, Pb often contains $^{210}$Pb, which has a half-life of T$_{1/2}$=22.3\,yr and several radioactive daughter isotopes. There are several potential sources for this contamination \cite{heusser1995}. The isotope is part of the natural decay chain of $^{238}$U and is thus potentially already embedded into the ores. Moreover, during the fabrication process, $^{210}$Pb can be introduced into the bricks e.g. by contaminated charcoal during the reduction process in the furnace or the decay of Rn in the surrounding air.    
The only way to remove the contamination from a final Pb component is to wait for it to decay. Therefore, ancient Pb is desirable for applications close to low background detectors. Among the lowest observed contaminations is archaeological Roman Pb with 715\,$\mu$Bq~kg$^{-1}$ \cite{romanpb}. For standard Pb bricks specific activities up to the order of 100\,Bq~kg$^{-1}$ are usually found.

The $^{210}$Pb isotope itself decays mainly by the emission of $\gamma$-rays and Auger electrons to the excited (80.2\%) and ground state of $^{210}$Bi (19.8\%) via beta decay. The main signature line occurs at 46.5\,keV in the deexcitation of $^{210}$Bi. Meanwhile, the daughter isotope $^{210}$Bi (T$_{1/2}$=5.0\,d) almost exclusively contributes via a $\beta$-decay with an endpoint of 1.16\,MeV. The decay chain continues with the $\alpha$-emitter $^{210}$Po (T$_{1/2}$=138.4\,d) and ends in the stable $^{206}$Pb isotope.

\subsubsection{$^{210}$Pb within the inner shield layer: activity and MC}

The radiopurity of the Pb layers within the CONUS shield increases towards the detector chamber. For the floor of the chamber "LC2" Pb with a $^{210}$Pb activity of less than 0.2\,Bq~kg$^{-1}$ was used (for details on these Pb bricks see \cite{heusser1995}). 

The bricks making up the sides of the innermost Pb layer are a mix of low activity Pb, several decades old and used in previous experiments, and bricks refurbished from old Pb from the Freiburg Minster (see Figure \ref{innermostlayer_Pb_pic}). The top layer is less critical as the $^{210}$Pb decay radiation is suppressed by a Cu plate of 1\,cm thickness in-between.  

The activity of the side layers was determined in a dedicated measurement campaign including $\gamma$-screening measurements with the GIOVE detector \cite{giovepub} as well as $\alpha$- and $\beta$-counter measurements at Jagiellonian University, Krakow. For the melted Freiburg Minster Pb bricks selected in the final configuration of the CONUS shield a $^{210}$Pb activity of $<$1\,Bq~kg$^{-1}$ was determined. The activity of the Pb bricks from the other sources was found to be of the order of 2\,Bq~kg$^{-1}$ or lower. All in all, for the final inner shield configuration, the mean $^{210}$Pb contamination is small at the level of $<$1.7\,Bq~kg$^{-1}$.

As Figure \ref{fig:chapter2_mcefficiency} shows, low energetic radiation cannot penetrate the Cu cryostat. Therefore, for the Pb bricks only the decay of $^{210}$Bi is relevant. The $\beta$-decay electrons induce a {\it bremsstrahlung} continuum in the Pb. The decay was simulated in all inner shield layers, but Figure \ref{MC_innermostlayer_Pb} reveals that the side closest to the respective detector contributes by far the most. For the background model, the determined upper limits were used as conservative starting point and the spectra were re-scaled in context with the other respective contributions. Overall, a better agreement to the experimental data was found for a lower activity than the measured upper limits. Exemplary for C1 (with the largest contribution), in the CE$\nu$NS ROI within [0.4,1]\,keV$_{ee}$ 0.5\,d$^{-1}$kg$^{-1}$ are expected (measured overall background 5-15\,d$^{-1}$kg$^{-1}$), within [2,8]\,keV$_{ee}$ the predicted rate is 1.6\,d$^{-1}$kg$^{-1}$ (measured overall background 25-40\,d$^{-1}$kg$^{-1}$). At higher energies within [11,440]\,keV$_{ee}$ the rate is 140\,d$^{-1}$kg$^{-1}$ (measured overall background 400-800\,d$^{-1}$kg$^{-1}$).

\subsubsection{$^{210}$Pb within the cryostat end cap: activity and MC}

Next to the shield layers, $^{210}$Pb contaminations within the cryostats are also possible. Especially Pb-based soldering wire can exhibit high activities \cite{heusser1995}. Therefore, special attention was paid to keep the used amount as low as possible \cite{detectorpub}. Moreover, surface contaminations on Cu parts or the diode itself from working spaces or surrounding air with a high Rn concentration in a poorly ventilated environment are possible. 

For contaminations within the cryostat end cap it is possible to observe the 46.5\,keV line of $^{210}$Pb itself\footnote{The shape of the line is not fully Gaussian, especially for C1. This might indicate that the decay occurs on the surface of the diode, where the charge collection efficiency could be diminished.}. The line is indeed visible in all four CONUS detectors with count rates of (5-10)\,d$^{-1}$. 

\begin{figure}[h] 
    \centering
\includegraphics[width=0.47\textwidth]{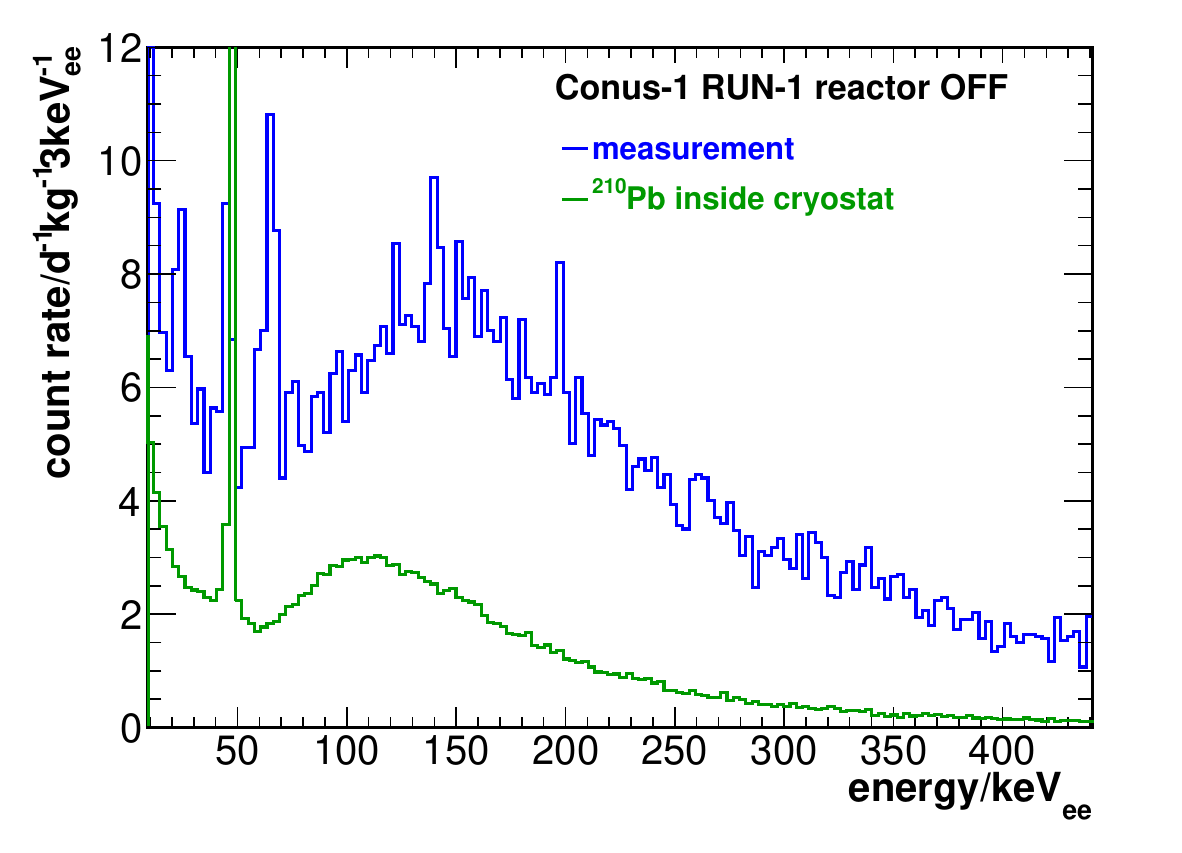}
	\caption{Comparison of the MC of $^{210}$Pb decays within the cryostat to the data of C1 RUN-1. For the other detectors, this contribution is smaller. It becomes evident that the 46.5\,keV line in the data is induced by these decays. 
	}		
	\label{210Pb_MC_withincryostat}
 \end{figure}

\begin{figure}[h] 
    \centering
\includegraphics[width=0.4\textwidth]{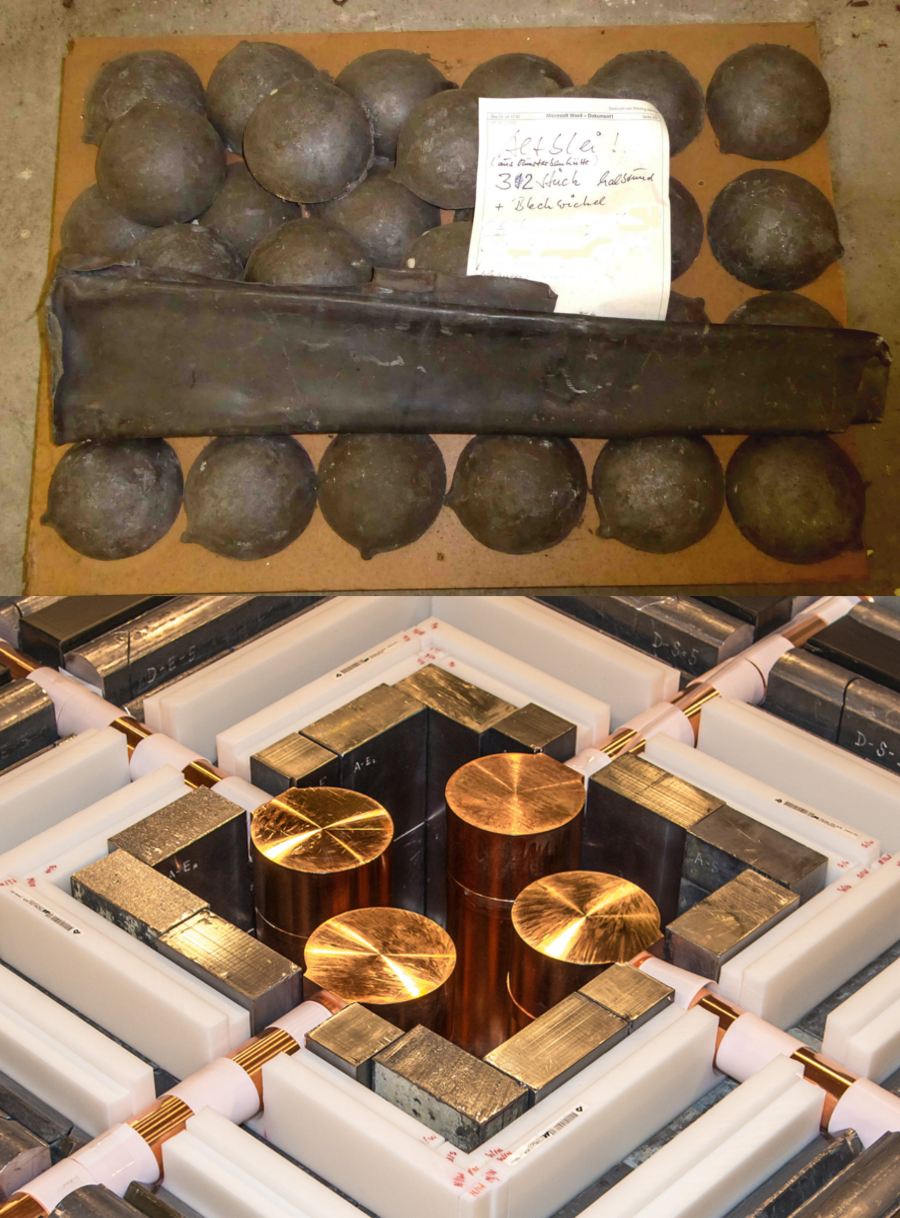}
	\caption{Top: Old Pb half-spheres from the Freiburg Minster, low in $^{210}$Pb activity; Bottom: Innermost Pb layer of the CONUS shield consisting among others of the Pb bricks melted from the Freiburg Minster Pb.
	}		
	\label{innermostlayer_Pb_pic}
 \end{figure}

\begin{figure}[h] 
    \centering
	\includegraphics[width=0.5\textwidth]{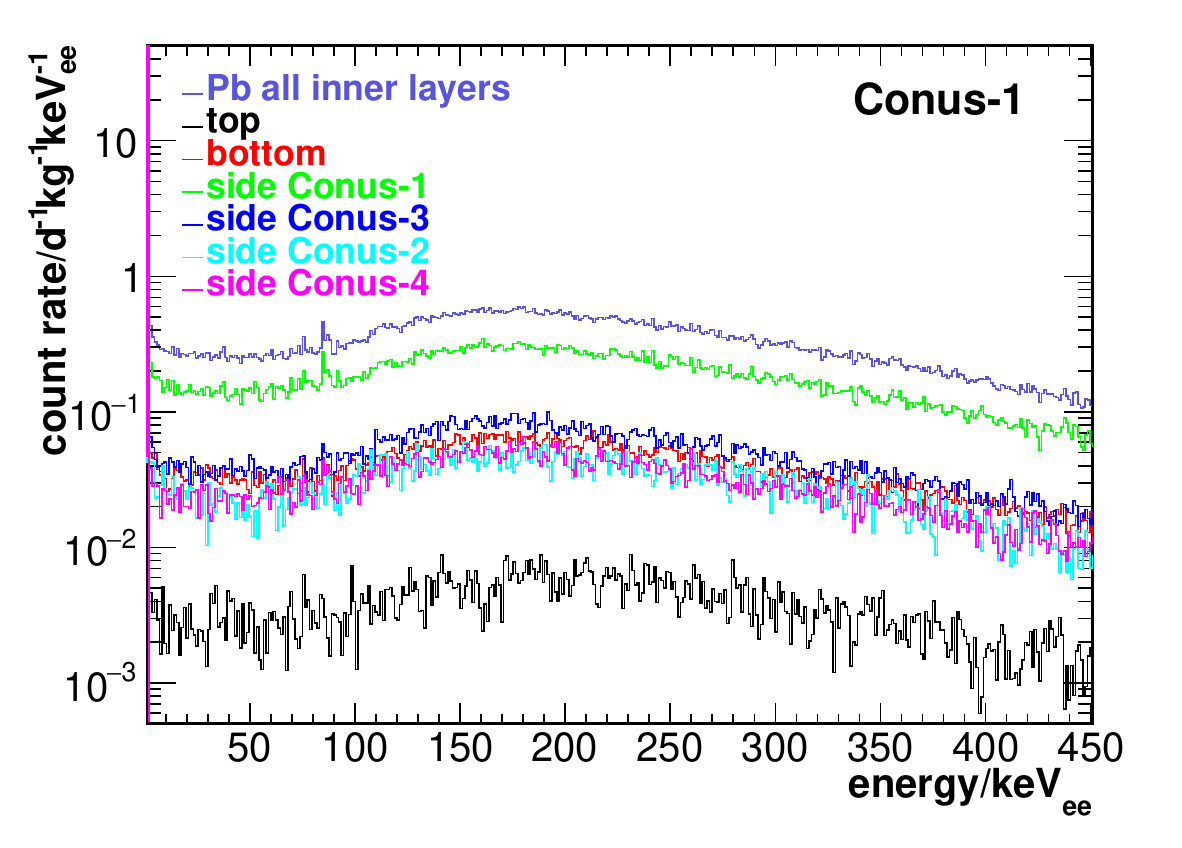}	
	\caption{MC simulation of $^{210}$Pb decays within the innermost Pb layer of the CONUS shield registered by C1, scaled with the respective measured activities of the Pb bricks. It becomes evident that the contaminations within the bricks adjoined to the detector are the most relevant.}		
	\label{MC_innermostlayer_Pb}
\end{figure}

In the MC simulation of this contribution, not only $^{210}$Pb and $^{210}$Bi, but also $^{210}$Po decays are considered on the surfaces of the Cu parts around the diode and on the surface of the diode itself. Each isotope is simulated separately. While the dead layer with a thickness of $\mathcal{O}$(1\,mm) already completely shields against the $\alpha$-particles, this is not true for the passivation layer on one side of the diode with a thickness of $\mathcal{O}$(100\,nm) (see Section \ref{chapter3} for a description of the diode). The passivation only suppresses the signatures of the recoil of the nuclides emitting $\alpha$-particles. The exact thickness of the passivation layer is not known and the parameter was therefore varied in the post-processing of the MC from 25\,nm to 150\,nm. For C1, the best agreement with the spectral shape of the measured data is found for 25\,nm, while for the other detectors it is 75\,nm. The change in the spectral shape due to the variation of the passivation layer thickness is considered as a systematic uncertainty of the background model (as applied in \cite{bsmpaper}). The absolute contribution to the background model strongly varies between the different detectors as summarized in Table \ref{tab:bkgmodeltable}. It is the dominant contribution in the ROI for C1 and C3, while for C2 it is smaller than the muon-induced background. From the background model, $^{210}$Pb activities of 10-300\,$\mu$Bqcm$^{-2}$ for the diode surface and 0.5-6\,mBqcm$^{-2}$ for the Cu parts were derived. 
This component differs between the detectors with C1 having the largest contribution with 6.2\,d$^{-1}$kg$^{-1}$ in the CE$\nu$NS ROI within [0.4,1]\,keV$_{ee}$ (measured overall background 5-15\,d$^{-1}$kg$^{-1}$). Within [2,8]\,keV$_{ee}$ the predicted rate is 21.2\,d$^{-1}$kg$^{-1}$ (measured overall background 25-40\,d$^{-1}$kg$^{-1}$) and at higher energies within [11,440]\,keV$_{ee}$ the rate is 209\,d$^{-1}$kg$^{-1}$ (measured overall background 400-800\,d$^{-1}$kg$^{-1}$).
The spectrum of this contribution in the highE range including the 46.5\,keV line is shown in Fig. \ref{210Pb_MC_withincryostat}.

\subsection{Air-borne radon}
\label{sec:airbornern}
Rn belongs to the natural radioactivity decay chains of U and Th. While the parent isotopes decay within the material, the inert Rn can outgas to the surroundings due to its volatile nature \cite{heusser1995}. It is therefore found in the ambient air and can accumulate in enclosed spaces to up to several hundreds of Bqm$^{-3}$, especially if there is no fresh air ventilation present in surroundings of concrete or rock. The daughter isotopes tend to stick to cold surfaces and dust particles as well, which can lead to activity clusters. 

Inside A-408 at KBR, the Rn activity outside the shield is monitored constantly over time. While at the LLL at MPIK a mean activity of $\sim$60\,Bqm$^{-3}$ persists, within the enclosed space of the safety containment at A-408 more than 100\,Bqm$^{-3}$ are typically measured. Large fluctuations over time are observed induced by the temperature variations inside the building as well as changes in the air ventilation system especially during reactor outages. For various time periods correlations to the room temperature are observed.   

Regarding the background contribution to the Ge data, the isotope $^{222}$Rn with a half-life of 3.8\,d and a natural abundance of 90\% is more relevant, while $^{220}$Rn with a shorter half-life of 55.6\,s and a natural abundance of 9\% is less likely to reach the detectors inside the massive CONUS shield. To suppress Rn background inside the detector chamber, a constant flushing with Rn free gas is required. The steel cage around the shield sealed with silicone is not tight enough to ensure a Rn free detector chamber. After a stop of the flushing, it takes about one hour for Rn to diffuse inside the chamber and to become apparent in the measured Ge data (see Section \ref{sec:rnatkbr}).

\subsubsection{Radon mitigation at KBR} 
\label{sec:rnatkbr}
Traditionally, Ge spectrometers are cooled with liquid nitrogen. The boil-off from the nitrogen storage tank (dewar) can be used to flush the surrounding of the detectors to keep away Rn. For CONUS, this is not possible as the detectors have to be electrically cooled for reactor safety reasons. As the best alternative, air from breathing air bottles from the KBR fire brigade department, that had been stored for at least two weeks for $^{222}$Rn to decay away, is utilized instead.

The Rn level around the detectors is monitored indirectly by looking for the signature of the Rn decays in the Ge data. The integral background in [20,440]\,keV$_{ee}$ as well as the line count rate in the 351.9\,keV line from the decay of the $^{222}$Rn daughter nuclide $^{214}$Pb is constantly evaluated with the one detector collecting data in the highE range.

\begin{figure*}[h] 
    \centering
    \includegraphics[width=0.75\textwidth]{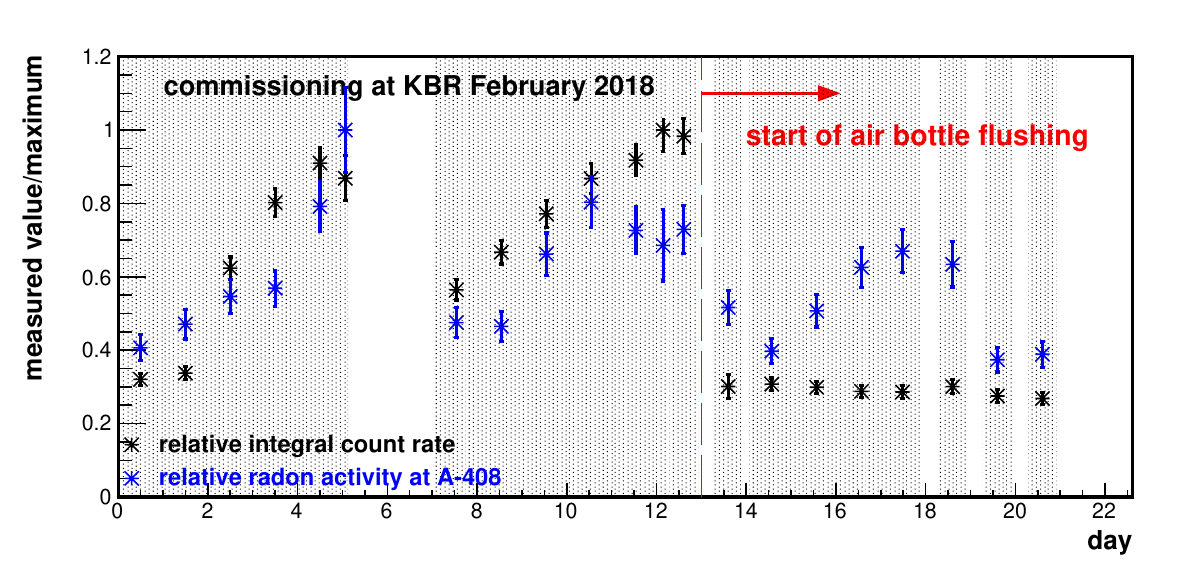}
    \caption{The relative development of the Rn activity over time in comparison to the relative integral count rate of C1 in [20,440]\,keV$_{ee}$ is shown during the commissioning at KBR before the beginning of RUN-1 in 2018. Before the start of the flushing with air from breathing air bottles, that have been stored long enough for the Rn to decay away, a clear correlation is seen. During the flushing a low and constant background count rate is measured.  }
    \label{fig:chapter4_startflushing}
\end{figure*}

Figure \ref{fig:chapter4_startflushing} shows how the integral count rate significantly drops with the start of the flushing, such that the remaining background is not dominated by Rn anymore. With the updated flushing system operational since the middle of June 2019 (outage of RUN-2) a complete suppression of the Rn background including the respective decay lines has been achieved. Before, remnants of Rn are seen in the data from a too weak air flushing, air bottles that have not been stored for long enough, and diffusion through the calibration source tube close to C3 (see Figure \ref{fig:chapter2_conusshield}). The latter is a PE tube used to insert from outside a $^{228}$Th wire source for calibration inside the shield. All in all, the air flux is regulated in such a way to guarantee a Rn free environment inside the detector chamber, while maintaining a reasonable bottle exchange rate. Before the optimization of the flushing, the strongest impact on the background spectra of the four detectors is observed for C3 with up to $\sim$20\,d$^{-1}$ in the 351.9\,keV line on some days. Thus, the Rn contributions need to be taken into account for the background model for RUN-1 and RUN-2 for each detector individually.  

\subsubsection{MC of radon background}

\begin{figure}[h] 
    \centering
    \includegraphics[width=0.5\textwidth]{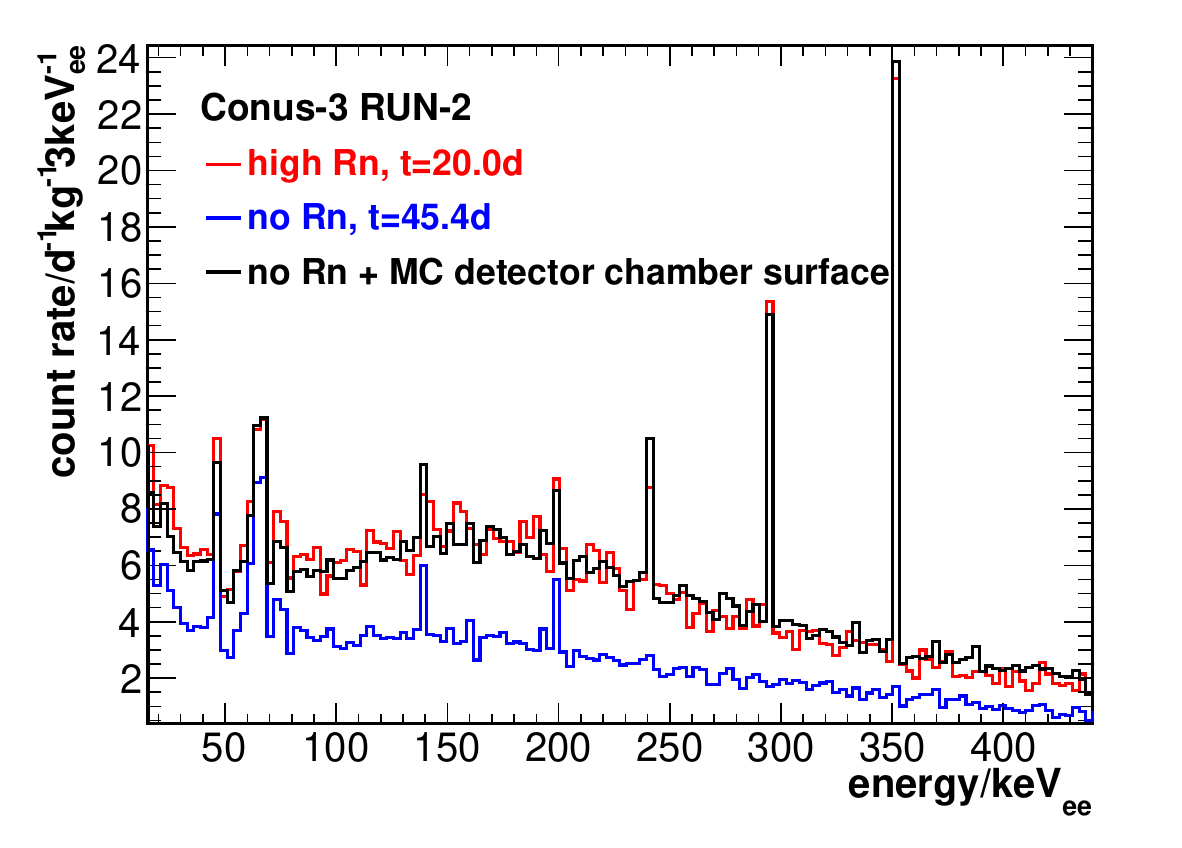}
    \caption{Simulation of Rn contribution to the background spectrum of C3: the MC simulation of the decay of the relevant Rn daughters on the surfaces within the detector chamber is added to a Rn free background spectrum. The MC is scaled such as to reproduce the data collected during times where a strong Rn background was present within the detector chamber. In this way, the contribution of Rn decays to the ROI can be estimated.}
    \label{fig:chapter4_rn_mc_detectorchamber}
\end{figure}

To study the Rn contribution to the background with the help of the MC simulation, two data sets from RUN-2 were selected (see Figure \ref{fig:chapter4_rn_mc_detectorchamber}). In one data set, $\sim$45\,kg$\cdot$d of exposure, during which no background lines from Rn was observed, were combined. Contrary, for the other data set the $\sim$20\,kg$\cdot$d of exposure with the strongest Rn contribution was chosen. 
The difference between the data sets is described by the MC simulation of Rn. The output of the simulations is added to the data spectrum without Rn contamination (blue) in Figure \ref{fig:chapter4_rn_mc_detectorchamber} to reproduce the data spectrum containing Rn (black). Only the decay chain of $^{222}$Rn within the detector chamber is considered. The contribution of the $\alpha$-decays of $^{222}$Rn and the polonium (Po) progenies are completely shielded by the Cu cryostat and can thus be neglected. Just $^{214}$Bi (T$_{1/2}$=19.9\,min) and $^{214}$Pb (T$_{1/2}$=26.8\,min) are simulated. The relevant part of the decay chain ends with $^{210}$Pb with a much longer half-life of 22.3\,yr. Within the highE range of the CONUS detectors, peaks are visible for $^{214}$Pb, while the decay of $^{214}$Bi contributes to the continuum. 
Consequently, $^{214}$Pb is scaled accordingly to the observed line count rates and $^{214}$Bi is scaled in a way to account for the missing contribution to the continuum.
In secular equilibrium, an equal activity of both isotopes is expected. However, due to the volatile nature of the Rn, the air flow of the flushing and temperature imbalances between the detector chamber and outside, the same activity for both isotopes is not necessarily expected \cite{radon_ineq1}, \cite{radon_ineq2}.

Three different locations for the decays were tested: the volume of the detector chamber, the surfaces within the detector chamber and the PE tube above C3 used for the insertion of the $^{228}$Th source. 
For the latter the activity per m$^{3}$ necessary to reproduce the spectrum is found to be by far too high to be realistic considering the tiny inner volume of the tube. This strongly indicates that the activity is not confined to the tube, but distributed within the chamber by outgassing from the plastic pipe over time. Moreover, especially towards lower energies a slightly better agreement for the decaying Rn progenies sticking to the detector chamber is found compared to the decays occurring homogeneously within the air volume of the detector chamber. The scaling of the MC spectra results in an activity ratio of $^{214}$Bi/$^{214}$Pb=1.5 for the two isotopes. Due to the better agreement with the data, we decided to use the simulation on the surfaces within the detector chamber as input for the background model.

Finally, according to the simulation for a count rate of 1\,d$^{-1}$ measured in the 351.9\,keV line, an integral contribution of 0.04\,d$^{-1}$kg$^{-1}$ in [0.4,1]\,keV$_{ee}$, 0.2\,d$^{-1}$kg$^{-1}$ in [2,8]\,keV$_{ee}$ and 18\,d$^{-1}$kg$^{-1}$ in [11,440]\,keV$_{ee}$ is expected. 
With observed line count rates up to $\sim$20\,d$^{-1}$, this background contribution is relevant for background modeling. The stability of the contribution for the different detectors and runs is discussed in detail in Section \ref{chapter6}.

 \subsection{Contaminations of components within cryostat}
\label{sec:contaminations}
All components within the cryostat as displayed in Figure \ref{fig:chapter2_conusshield}, including structure elements, electronics and cables, were tested on their radioactive contaminations in material screening measurements at LLL and, if a higher sensitivity was required, at Laboratori Nazionali del Gran Sasso (LNGS), Italy. If necessary, adaptions were made to find materials that meet the strict radiopurity criteria for the parts close to the diode. For all components activities as low as several $\mu$Bq and at maximum several 100\,$\mu$Bq per piece were achieved. A detailed list of the screening results is published in Table 3 in \cite{detectorpub}.

In the MC, the respective decay chains of natural radioactivity were simulated within the volume of the detector components. The screening results corresponding to a measured activity were used for normalization. All in all, less than 0.1\,d$^{-1}$kg$^{-1}$ are found in [0.4,1]\,keV$_{ee}$, while in [2,8]\,keV$_{ee}$ there are $\sim$0.5\,d$^{-1}$kg$^{-1}$ and for the highE range within [11,440]\,keV$_{ee}$ overall $\sim$7\,d$^{-1}$kg$^{-1}$ are expected. This is in agreement with the currently available highE background measurements, where no clearly visible background lines accounted to natural radioactivity are observed.   
Moreover, as depicted in Section \ref{chapter6} other background contributions were identified to make up most of the continuum, especially at low energies within the ROIs. They are mostly related directly or indirectly to the shallow depth location of the experiment.

 \subsection{Reactor-correlated background}

The reactor-correlated background inside A-408 at KBR was examined in a dedicated measurement campaign \cite{neutronpub}. Neutrons with energies of several MeV are created by the fission reactions inside the fuel elements within the reactor core at a distance of 17.1\,m to the experiment. The reactor-correlated fluence rate in the Bonner Sphere measurements in A-408 was found to be about two times smaller than at Earth's surface (energy range from thermal energies around 25\,meV up to 10\,GeV). About 80\% of the neutrons arriving at the exact location of the CONUS shield are thermal and the maximum neutron energy lies below 1\,MeV. Inhomogeneities in the spatial distribution of the thermal neutron fluence rate of $\sim$20\% were found. The fission neutron fluence rate from the reactor core is suppressed by about twenty orders of magnitude by the borated water inside the reactor core, the steel of the reactor pressure vessel and the steel-enforced concrete of the biological shield and the walls around A-408. It was shown with the help of MC simulations including the full measured spectrum from thermal energies to the maximum energy below 1\,MeV that the contribution within the CONUS shield is $\sim$0.01\,d$^{-1}$kg$^{-1}$ in [0.4,1]\,keV$_{ee}$ and $\sim$0.02\,d$^{-1}$kg$^{-1}$ in [1,10]\,keV$_{ee}$. This count rate is considered negligible in comparison to the other background contributions. 

The $\gamma$-ray background was characterized by placing the CONRAD HPGe spectrometer without any shield inside A-408 at various places. Above the range of natural radioactivity $\gamma$-lines from the decay of $^{16}$N inside the cooling cycle of the reactor as well as $\gamma$-lines from neutron captures inside the steel enforced concrete walls were observed. Within the ROI for CE$\nu$NS with the help of MC simulations this background was found to  also be negligible inside the CONUS detector chamber.

\subsection{Spent-fuel elements induced background}

The overburden directly above A-408 consists partially of the spent fuel storage pool as well as the pool, where the spent fuel elements are placed into containers for transportation. At maximum, about 133 spent fuel elements can be located directly above A-408. The spent fuel elements produce neutrinos, neutrons and $\gamma$-rays. 
The maximum neutrino energies emitted by the elements are all below 5.5\,MeV an hour after discharge from the reactor core \cite{spentfuelneutrinos}, which translates to a recoil energy below 0.16\,keV$_{ee}$. This is well below the energy threshold of the CONUS detectors \cite{cevnspub}.
For neutrons, each fuel element emits about 10$^{9}$ neutrons per second in the first year after being removed from the reactor core \cite{spentfuelneutronflux1}, \cite{spentfuelneutronflux2}. The main neutron source is $^{244}$Cm \cite{spentfuelneutronenergy}, which produces neutrons by spontaneous fission, that follows a Watt distribution with an average energy of $\sim$2\,MeV. This is used as an input for a MC simulation, where the neutrons are propagated through the holder of the spent fuel elements, the borated water, the steel-lining of the pool and the concrete structure of the building. In this way, an upper limit of the neutron fluence rate of $<$1\,cm$^{-2}$d$^{-1}$ inside A-408 can be set, which is negligible in comparison to the measured neutron fluence rate during the Bonner Sphere measurement for the site characterization \cite{neutronpub}.
Any impact of $\gamma$-ray background from the spent fuel elements is excluded by the measurements with the CONRAD HPGe spectrometer. Measurements behind the CONUS shield on the site opposite to the direction of the cooling cycle pipes during reactor ON periods as well as measurements during an outage resulted in background spectra comparable to the data collected at the LLL at MPIK. All in all, no influence of the spent fuel elements on the CONUS background is expected.

\section{Background Model}
\label{chapter6}
 
In the background model, all previously introduced contributions are summed up. The contributions are scaled according to all available information such as activities or line count rates in the background data (see Section \ref{chapter4}). This however still leaves degrees of freedom in the scaling of the different spectra. In the following, the proceeding in the adaption of the spectra to the data is described and the result, the background model below 10\,keV$_{ee}$, is compared to the data. A Kolmogorow-Smirnow-test \cite{kstest} is used to confirm the agreement.

For the prompt $\mu$-induced background within the highE range, where no physics analysis is carried out, the background data without $\mu$-veto is used instead of the MC due to the observed small discrepancies within [10,100]\,keV$_{ee}$. For the lowE range, the MC simulation of the component validated by the comparison to the data without $\mu$-veto is applied as described in Section \ref{subsec:muons} including the small adjustments of the electromagnetic and muon-induced contributions against each other to achieve an agreement with the data without $\mu$-veto. The spectra for both energy ranges are scaled with the $\mu$-veto suppression efficiency of 97\% resulting in an absolute count rate of $\sim$4\,d$^{-1}$kg$^{-1}$ within [0.4,1]\,keV$_{ee}$. Next, the $^{210}$Pb contribution from the innermost shield layer is added, using the upper limits and finite measurements as starting point for the normalization as described in Section \ref{subsec:pb}. To match the experimental data above 100\,keV$_{ee}$, for most detectors a down-scaling of the $^{210}$Pb contribution is necessary, which does not create tension as most of the activity measurements resulted in upper limits. The two background contributions are degenerated at higher energies above $\sim$200\,keV$_{ee}$, while the $\mu$-induced background contributes more at lower energies because of neutron-induced recoils.  

In a further step, the spectral distribution of the $\mu$-induced metastable Ge states are added, scaled according to the observed background line count rates.  
Moreover, the spectra from the decays of the cosmogenic isotopes are normalized to the observed background line count rates if available or to the averaged production rates from literature (for details see Section \ref{subsec:cosmicactivation}) and added as well.
The large uncertainties on the production rates result in minor uncertainties on the spectral shape of the background model, which nevertheless might be relevant depending on the physics topics under investigation.

The presence of the 46.5\,keV line indicates a $^{210}$Pb contribution within the cryostat end cap (see Section \ref{subsec:pb}). Contaminations of the diode surface as well as the surfaces of the Cu diode holder were simulated and scaled in such a way to describe simultaneously both the highE and lowE spectra. The required normalization factors are found to result in reasonable surface contamination activities. The scaling leads to a slight overestimation of the line count rate in the 46.5\,keV line. However, the precise location of the contamination is not known as well as the exact thickness of the passivation layer, which is highly relevant for this contribution. The shape uncertainties derived from a variation of the passivation layer thickness in the MC are used as systematic uncertainties in the likelihood fit of the BSM analysis within the [2,8]\,keV$_{ee}$ range of the CONUS data \cite{bsmpaper}.  
Other than $^{210}$Pb, there are no relevant contaminations inside the cryostat, which was assured by a careful selection of the materials beforehand in a screening campaign.

For the Rn contribution, the background line count rates observed in the highE range data (e.g. 351.9\,keV line of $^{214}$Pb), if available, were used as indication for the normalization. The assumption is made that the available data gives a reasonable representation of the mean Rn background over time, which can also provide the normalization for the time periods when no highE data is available for a detector. Fluctuations are considered negligible compared to the overall background level.

\begin{figure*}[h]
 
	\centering
	\includegraphics[width=10.3cm]{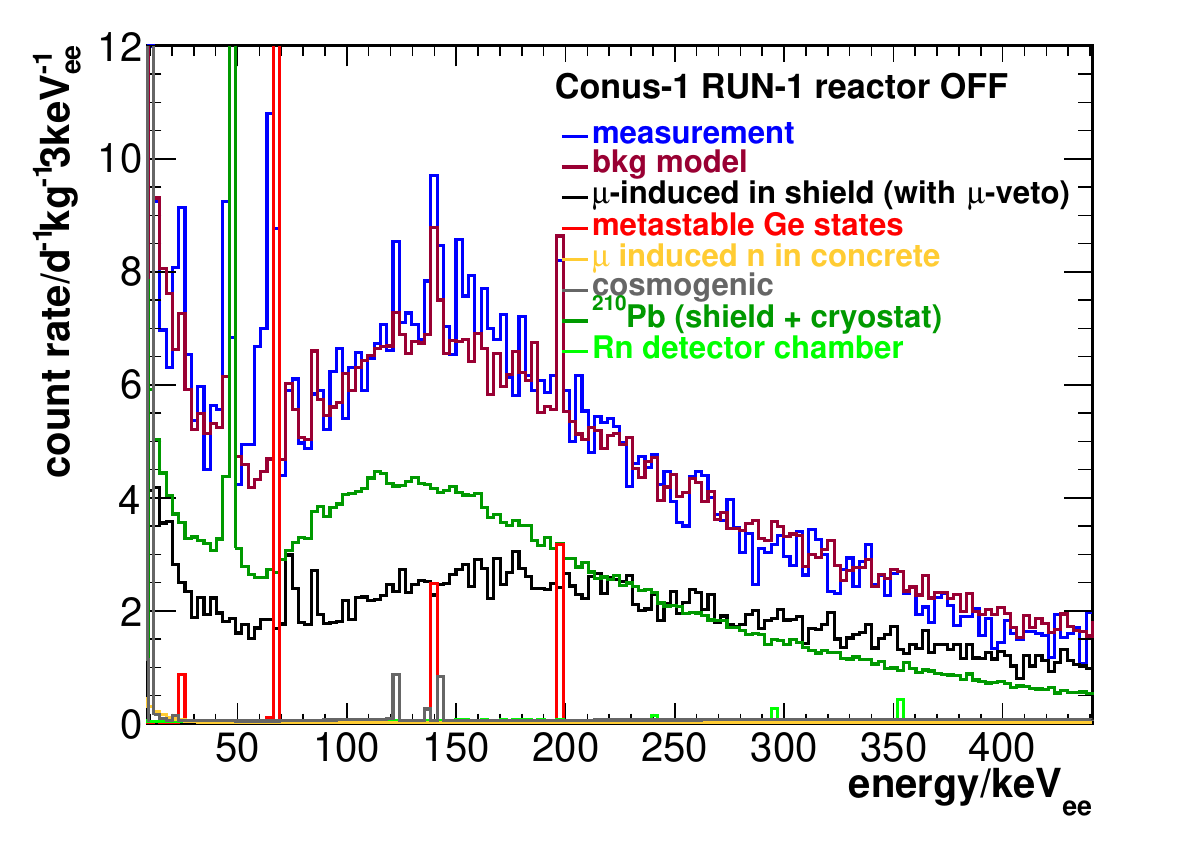}

	\centering
	\includegraphics[width=10.3cm]{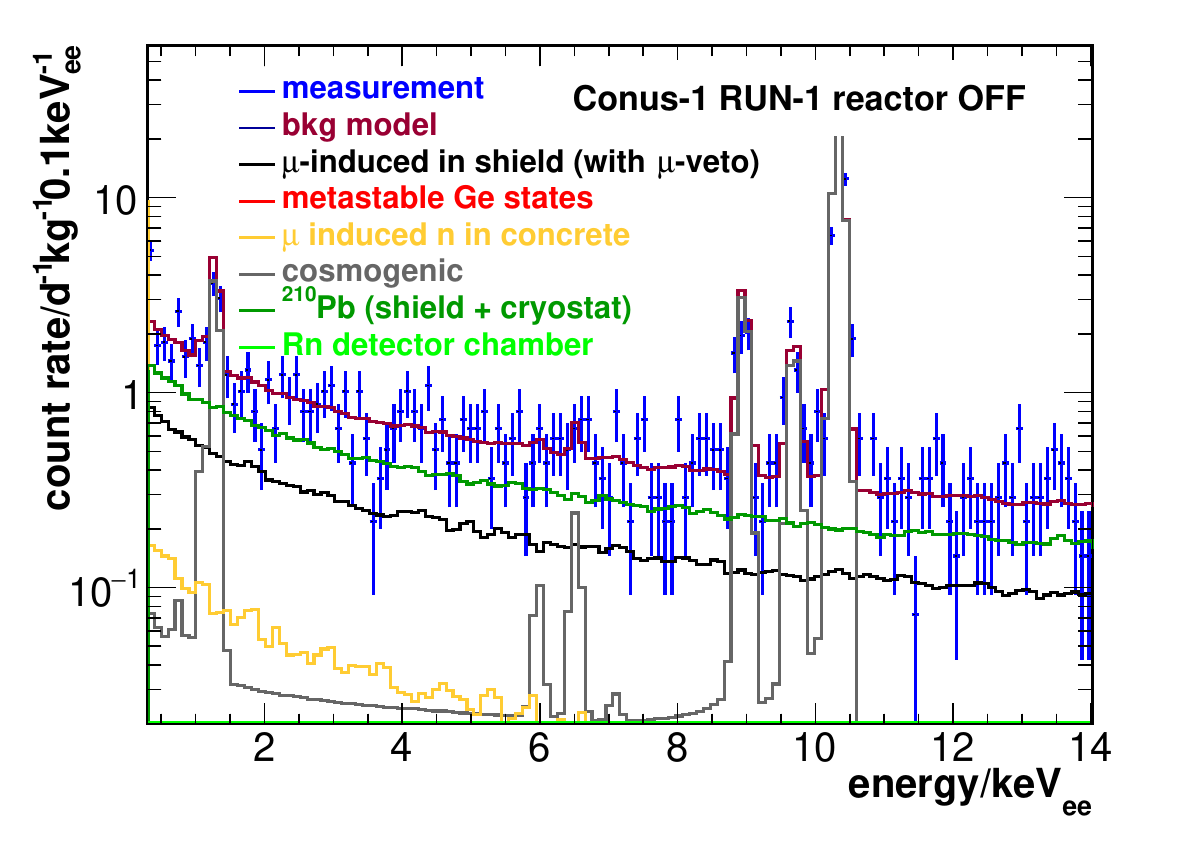}

	\centering
	\includegraphics[width=10.3cm]{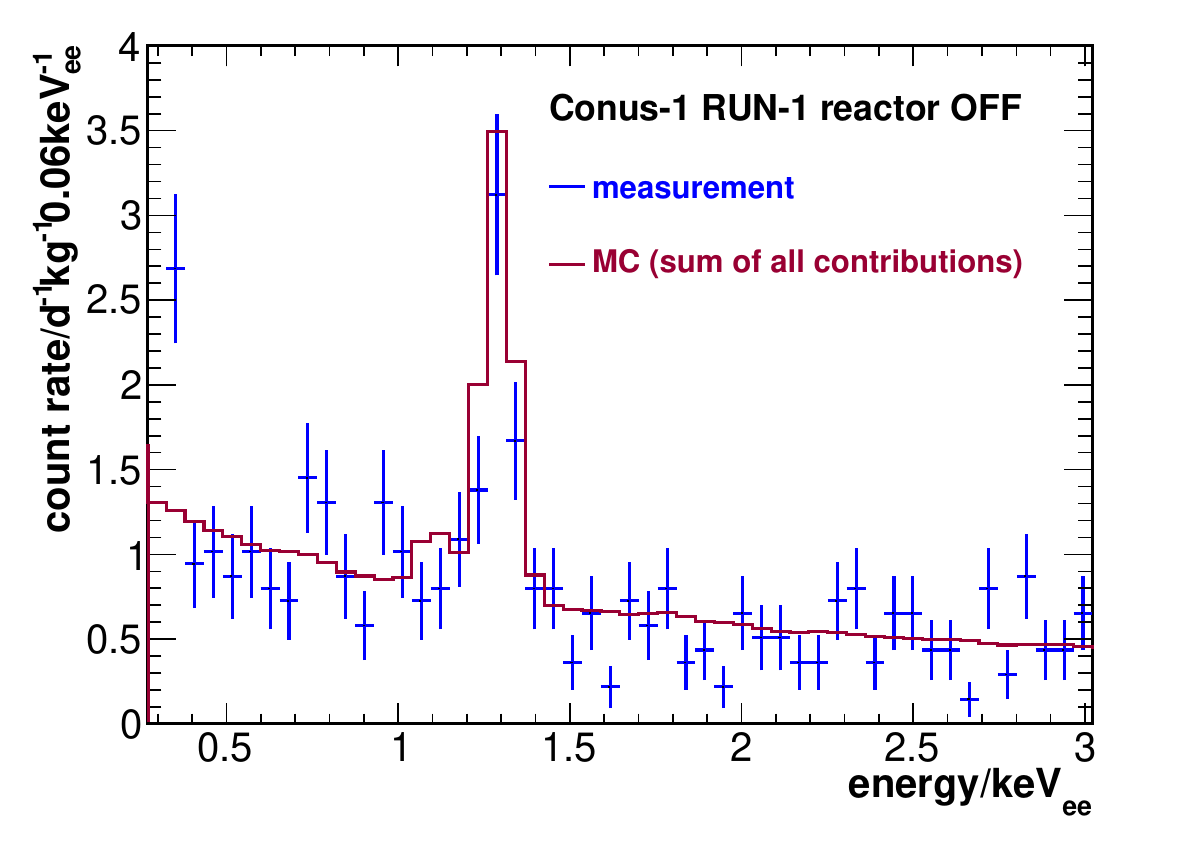}

	\caption{The measured background spectrum of C1 is depicted together with the background model in highE range as well as the lowE range and an even smaller energy range focusing on the ROI for CE$\nu$NS. The MC spectra from the various background sources are shown in different colors. The strong increase within the lowE range below 400\,eV$_{ee}$ is attributed to noise and is fitted with an exponential two parameter model.}
	\label{fig_completebkgmodelc1} 
\end{figure*}

\begin{figure*}[h]
	\centering
	\includegraphics[width=10.3cm]{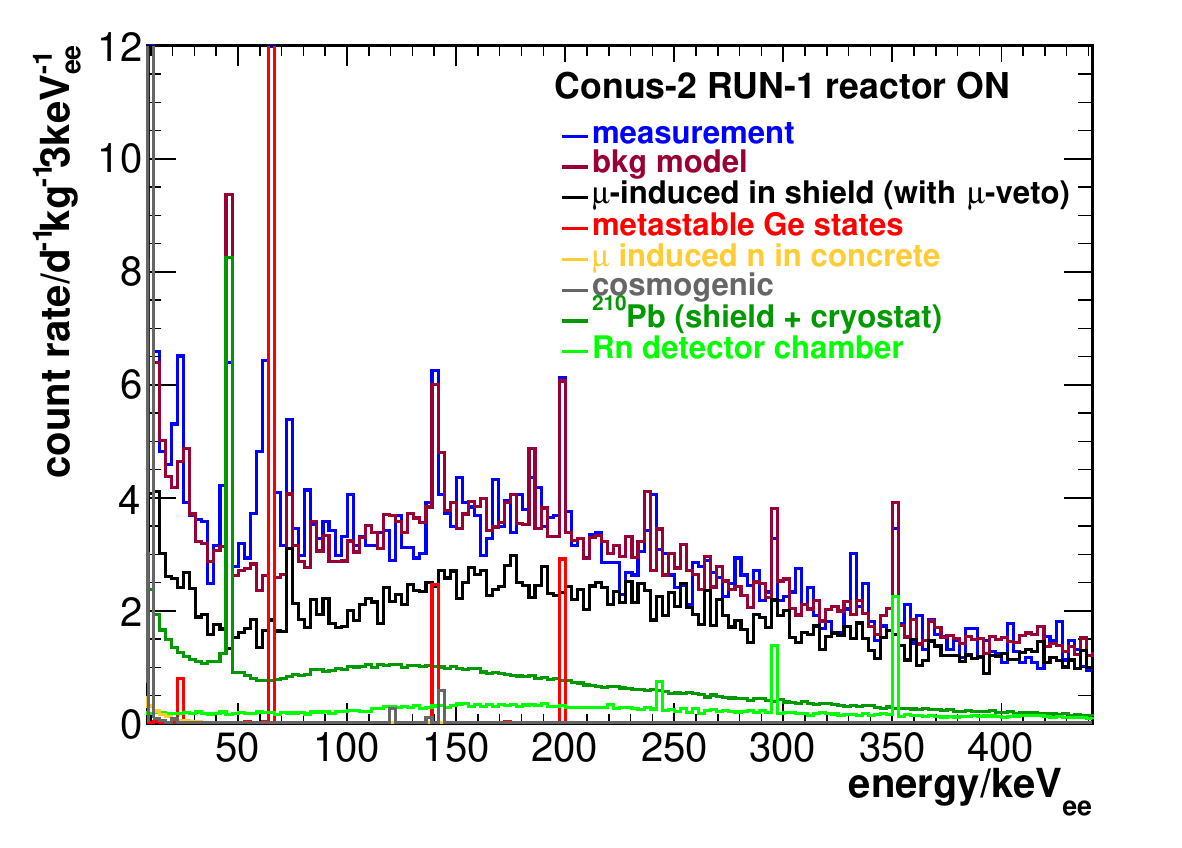}
	\label{fig:bkgmodelc2}
 
	\centering
	\includegraphics[width=10.3cm]{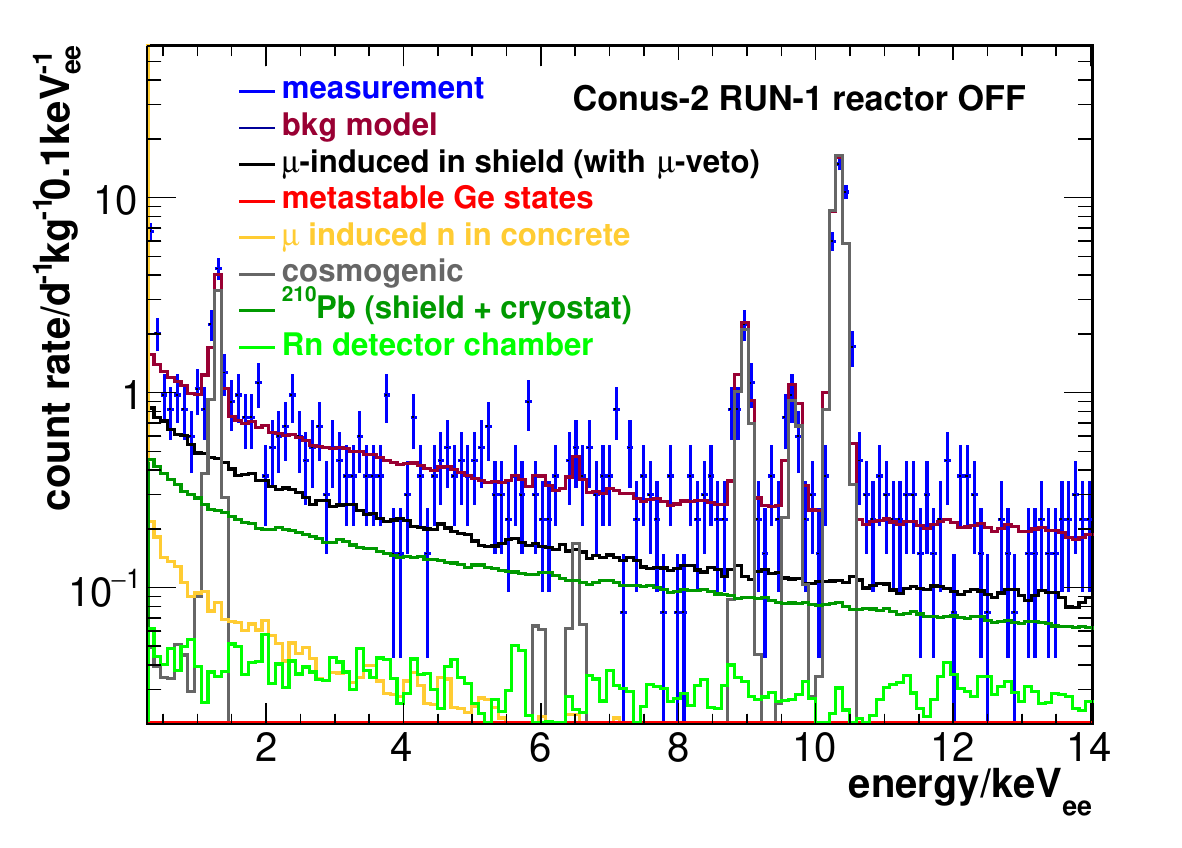}	
	\label{fig:bkgmodelc2low}
	
		\centering
	\includegraphics[width=10.3cm]{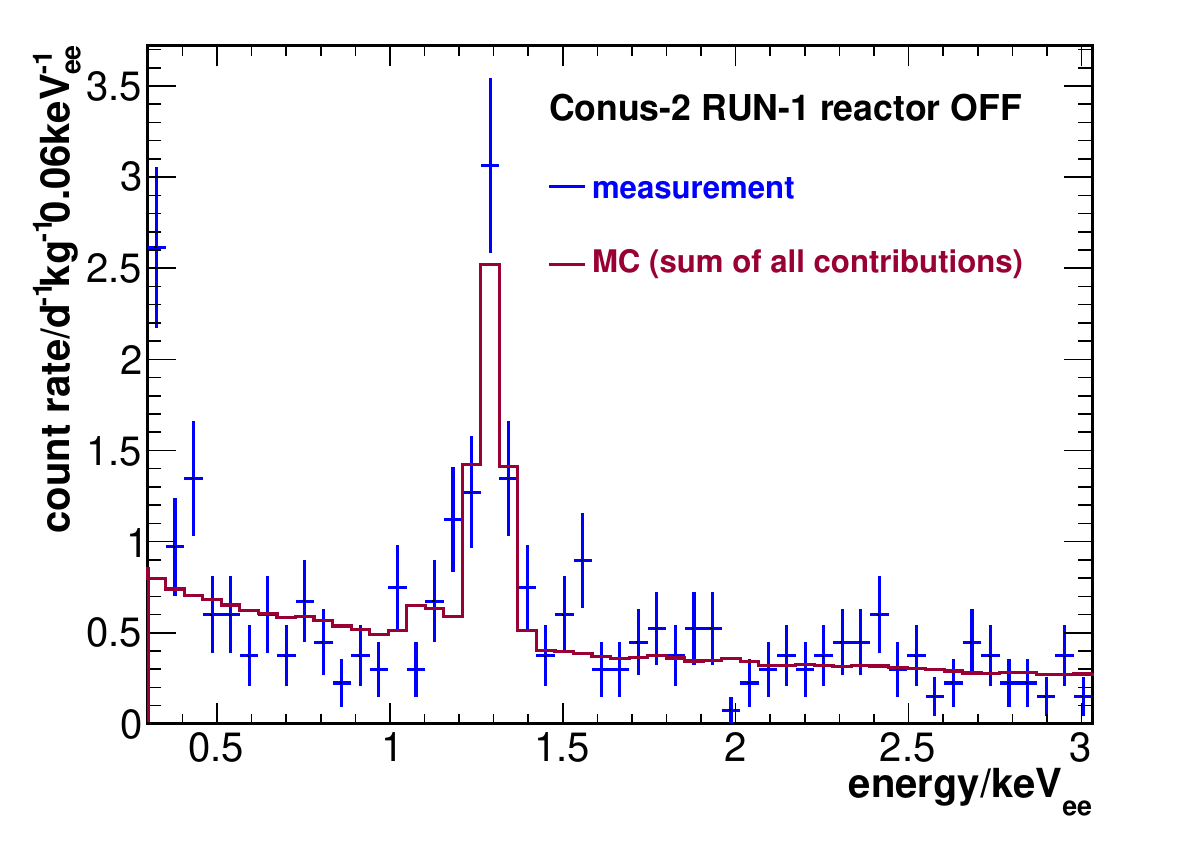}	
	\label{fig:bkgmodelc1lowzoom}
	
 	\caption{The measured background spectrum of C2 is depicted together with the background model in highE range as well as the lowE range and an even smaller energy range focusing on the ROI for CE$\nu$NS. The MC spectra from the various background sources are shown in different colors. The strong increase within the lowE range below 400\,eV$_{ee}$ is attributed to noise and is fitted with an exponential two parameter model. For the highE range, reactor ON data were used, as no reactor OFF data is available.}
	\label{fig_completebkgmodelc2}
\end{figure*}

\begin{figure*}[h]
	\centering
	\includegraphics[width=10.3cm]{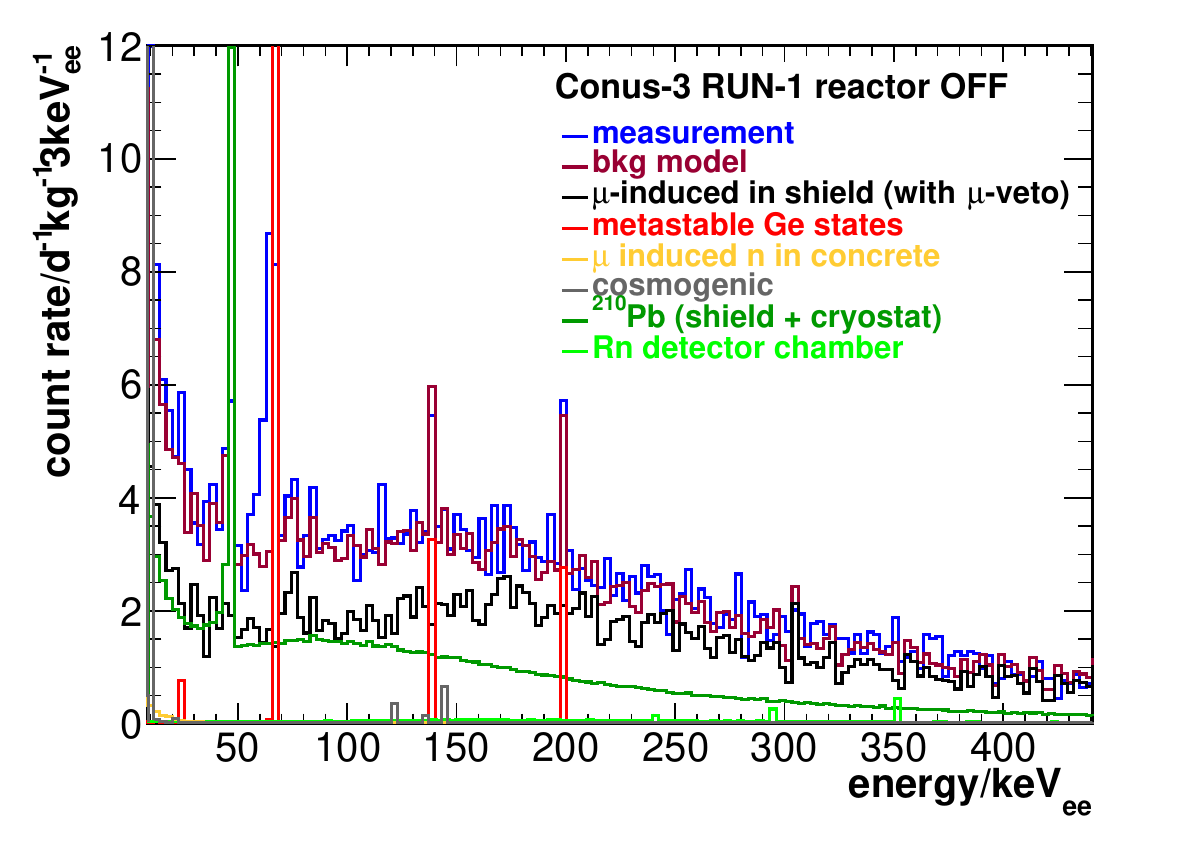}

	\centering
	\includegraphics[width=10.3cm]{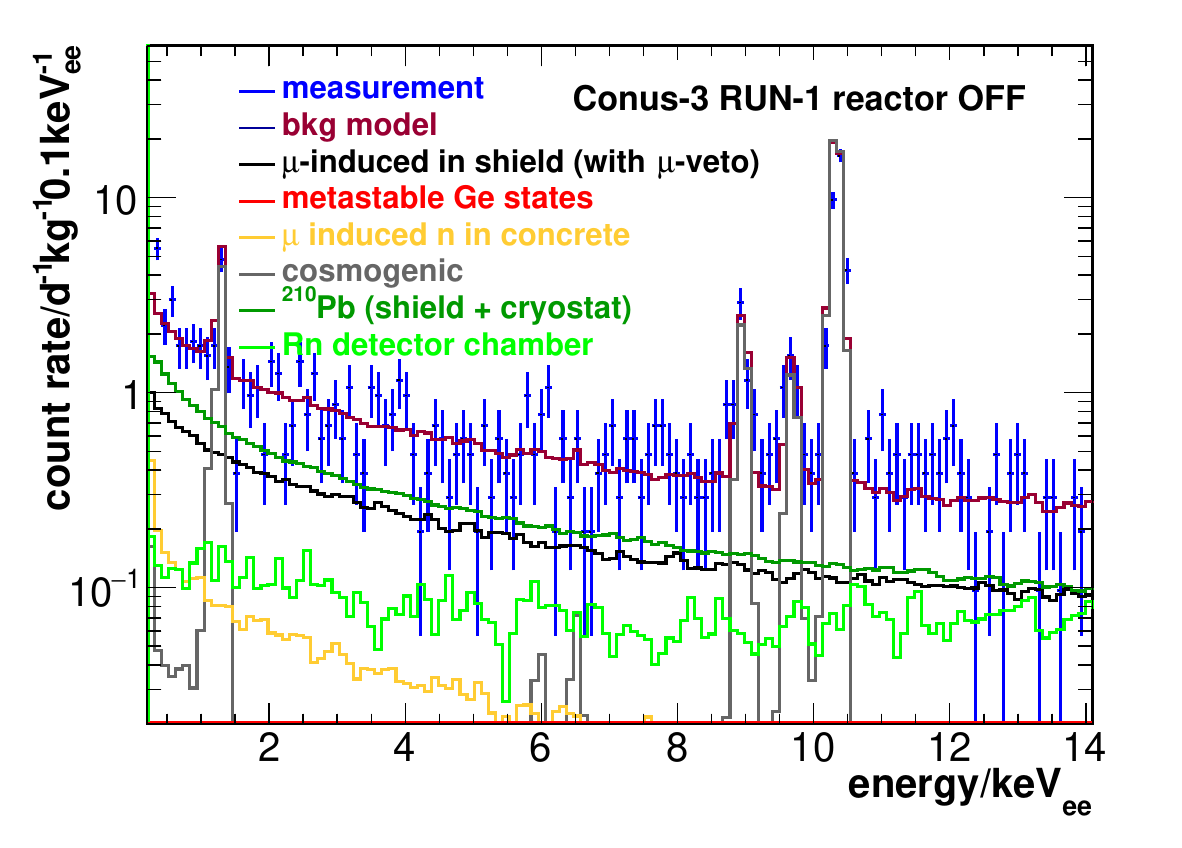}

     \centering
	\includegraphics[width=10.3cm]{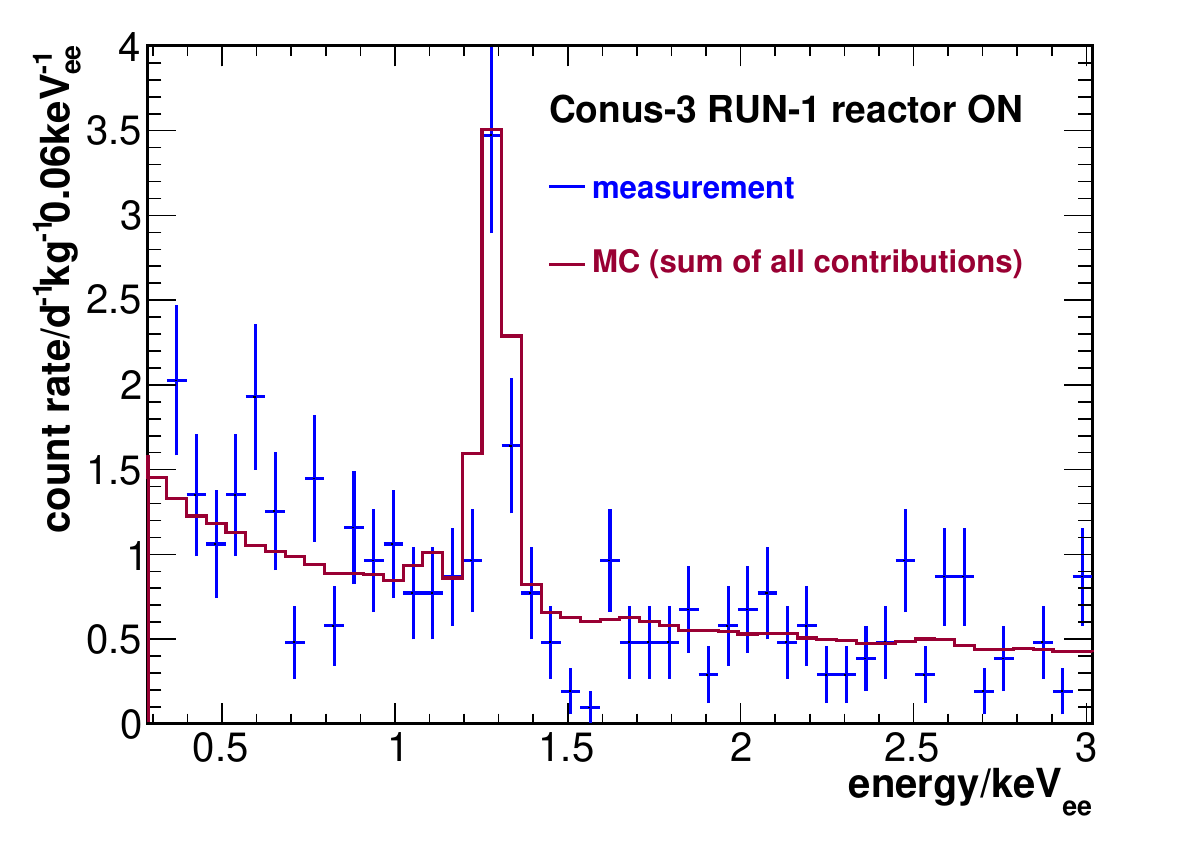}

	\caption{The measured background spectrum of C3 is depicted together with the background model in highE range as well as the lowE range and an even smaller energy range focusing on the ROI for CE$\nu$NS. The MC spectra from the various background sources are shown in different colors. The strong increase within the lowE range below 400\,eV$_{ee}$ is attributed to noise and is fitted with an exponential two parameter model. For the highE range, reactor ON data were used, where the Rn background was completely suppressed.}
	\label{fig_completebkgmodelc3}
\end{figure*}

The full background model decomposition and a comparison to the experimental data in various energy ranges is shown for C1 to C3 in Table \ref{tab:bkgmodeltable}. The table also includes a separation into the different background components. If possible, reactor OFF data are selected for the comparison. But due to the limitations of the DAQ this is not always possible for the highE range (see Section \ref{subsec:dataforbkgmodel}).
In the Figures \ref{fig_completebkgmodelc1} to \ref{fig_completebkgmodelc3}, the corresponding background spectra for the lowE and highE ranges are displayed in comparison to the MC simulation for the three detectors.

Within the highE range above 100\,keV$_{ee}$ the continuum is dominated by the residual prompt $\mu$-induced background with a contribution of approximately 40-70\% to the total background, even after application of the $\mu$-veto. The remaining sizable background contribution (20-50\%) is related to $^{210}$Pb: the contaminations within the cryostat are relevant as well as decays on the surface of the diode and the Cu parts within the end cap. The latter varies between the different detectors, with C1, the first detector manufactured, observing the largest contribution. 

Towards energies below 10\,keV$_{ee}$, the $\mu$-induced background as well as the $^{210}$Pb contribution within the cryostat still dominate. As stated in Section \ref{subsec:muons}, at these energies recoils of $\mu$-induced neutrons make up most of the $\mu$-induced background. Next to the $\mu$-induced neutrons inside the shield also $\mu$-induced neutrons in the concrete of the walls are relevant. Moreover, the low energetic decays of the $^{210}$Pb decay chain either close to the diode or on its surface are highly relevant.
Furthermore, there is a small contribution of 2-5\% by the decay of cosmogenic-induced isotopes within the Ge and Cu. Next to the $\beta$-decay of $^{3}$H this continuum is created by the slow pulses of the $\gamma$-rays emitted in EC decays.  
 
To validate the background model, a Kolmogorow-Smirnow-test is carried out. The test allows for a comparison of binned data if the bin width is small compared to all physics effects. This is the case for the chosen 10\,eV$_{ee}$ binning, which is the minimal bin width used in the analyses, in comparison to the pulser resolution of (60-75)\,eV$_{ee}$ (in terms of full-width at half-maximum). It confirms that the background model is compatible with the reactor OFF data for the [0.4,1]\,keV$_{ee}$ and [2,8]\,keV$_{ee}$ energy range. Only for C1 RUN-1 a significant better agreement was found when restricting the test to [0.4,0.75]\,keV$_{ee}$ due to a small overfluctuation in the range of [0.75,1]\,keV$_{ee}$. Thus, in this case the upper end of the CE$\nu$NS ROI was lowered respectively \cite{cevnspub}. This does not influence the signal expectation, but only restrict the background range included within the fit.

Below $\sim$400\,eV$_{ee}$ (depending on the detector and run) the measured background data start to increase exponentially towards lower energies above the MC background model. The origin of this increase is due to noise that passes the trigger condition of the DAQ.  
The stability of the noise is independent of the stability of the background. It has been examined and described separately \cite{detectorpub}. In particular, an impact of changes in the room temperature via the performance of the electrically powered cryocooler can be observed. Large time periods of exposure are removed from the CE$\nu$NS analysis data sets due to this observation \cite{detectorpub}.   

The described background model is an input to the likelihood analyses looking for CE$\nu$NS or for BSM physics, with $\mathcal{L}_{i}$ representing the likelihood of detector and run $i$:
\begin{equation}
\begin{split}
\mathcal{L} &=\prod_{i}\mathcal{L}_{i}(ON)\cdot\mathcal{L}_{i}(OFF) \text{ with pull terms} \\  &=\prod_{i}\mathcal{L}(s_{i}+b_{i},\theta_{1,i},\theta_{2,i})\cdot\mathcal{L}(b_{i},\theta_{1,i},\theta_{2,i})\\
&\quad\text{with pull terms}
\end{split}
\end{equation}
In the framework of the likelihood function, the predicted CE$\nu$NS signal spectrum is scaled by the parameter $s$, while the parameter $b$ allows for a normalization of the background model to adapt to the data sets for each run. All detectors and runs are fit simultaneously.
For all fit results obtained so far, the $b$ parameter agrees with unity within one sigma showing that the background model can be used as a good description of the experimental data. The noise contribution is included separately in the likelihood model by an exponential function including two unconstrained fit parameters $\theta_{1}$ and $\theta_{2}$.
While the $b$ parameter is unconstrained as well, for the BSM analyses \cite{bsmpaper} at higher energies within [2,8]\,keV$_{ee}$, two additional parameters with pull terms are used to model small shape uncertainties with a polynomial function of second order. The pull terms on these nuisance parameters of the function ensure that the variation is restricted to $\leq 5\%$ per bin. The shape uncertainties result from the uncertainties on the production rate of the cosmogenic isotopes (see Section \ref{subsec:cosmicactivation}) as well as from uncertainties in the thickness of the passivation layer affecting the shape of the $^{210}$Pb decay spectrum on the surface of the diode (see Section \ref{subsec:pb}). Uncertainties on the muon-induced background contribution are not taken into account as its shape is fixed by the comparison to data without $\mu$-veto. Changes in the normalization need to be counter-balanced by the rescaling of other contributions and overall the impact on the spectral shape of such a rescaling is minor. The impact of these uncertainties on the upper limit of the CE$\nu$NS analysis derived in \cite{cevnspub} is negligible, as this analysis is restricted to a much smaller energy range. Further, uncertainties of this order are balanced by the other fit parameters without significant changes of the upper limit. 

\begin{table*}[btp]
\begin{center}
\begin{footnotesize}
\begin{tabular}{lccccc}
\hline

 energy ranges          & [0.4,1]\,keV$_{ee}$ & [2,8]\,keV$_{ee}$  & [40,48]\,keV$_{ee}$  & [10,100]\,keV$_{ee}$  & [100,440]\,keV$_{ee}$ \\ \hline
C1     &	RUN-1 OFF   &            RUN-1 OFF      &      RUN-1 OFF   &         RUN-1 OFF  &        RUN-1 OFF         \\ \hline
data [d$^{-1}$kg$^{-1}$]      &	  11.5$\pm$0.9 &  40.5$\pm$1.8        &   22.1$\pm$0.9      &    262$\pm$3 &    529$\pm$4  \\ 
bkg model [d$^{-1}$kg$^{-1}$]    &	  11.9    &    40.2               &   29.0              & 264          &     529   \\ \hline
prompt $\mu$-induced                &	    34\%    &    33\%              &   19\%             &   27\%         &  44\%    \\
metastable Ge states                &    0.4\%       &     0.3\%        &        0.3\%         &     7\%          &  1\% \\
$\mu$-induced n concrete              &	3\%       &      2\%                 &  0.1\%          &   0.3\          &     0.1\%\\
cosmogenic  &	3\%             &     5\%                      &    0.5\%              &    17\%                   &     2\%                   \\
$^{210}$Pb shield  &	        4\%     &      6\%   & 8\%                  &       11\%            &     27\%                                \\
$^{210}$Pb within cryostat &	    54\%         &       52\%                    &     70\%              &     35\%                  &     22\%                   \\
other cont. within cryostat   &	       1\%      &   1\%                        &       2\%           &          2\%             &       3\%                 \\ 
airborne Rn  &	      0.2\%           &        0.3\%                   &     0.4\%              &     0.5\%                  &           1\%             \\\hline

C2     &	    RUN-1 OFF           &         RUN-1 OFF                    &          RUN-1 ON$^{(*)}$         &         RUN-1 ON$^{(*)}$              &       RUN-1 ON$^{(*)}$                 \\ \hline
data [d$^{-1}$kg$^{-1}$]    &6.4$\pm$0.7   &    26.8$\pm$1.5      &      14.3$\pm$0.8    &   173$\pm$3 & 338$\pm$4         \\
bkg model [d$^{-1}$kg$^{-1}$]   &	    7.0         &               25.8            &                  16.1     &     169                  &     346                   \\ \hline
prompt $\mu$-induced &	    55\%         &       48\%                    &      28\%             &         39\%              &      65\%                 \\ 
metastable Ge states &	     0.6\%        &    0.3\%                       &      0.4\%             &  9\%                  &      2\%                  \\
$\mu$-induced n concrete &	5\%             &      3\%                     &     0.2\%             &     0.4\%                  &            0.2\%                  \\
cosmogenic  &	   3\%          &      5\%                     &     0.3\%             &     19\%                  &          1\%              \\
$^{210}$Pb shield  &	        2\%     &   2\%                       &        4\%           &      5\%                 &     11\%                   \\
$^{210}$Pb within cryostat &	   29\%          &          30\%                 &        60\%           &   20\%                    &     8\%                   \\
other cont. within cryostat   &	     1\%        &       4\%                    &     4\%              &         4\%              &        4\%                \\ 
airborne Rn  &	    4\%          &     8\%                      &  3\%                 &    4\%                   &     9\%                  \\\hline
C3     &	      RUN-1 OFF        &                  RUN-1 OFF           &           RUN-2 ON          &    RUN-2 ON                    &            RUN-2 ON              \\\hline
data [d$^{-1}$kg$^{-1}$]  &13.1$\pm$1.2      &      37.6$\pm$2.0  &    14.5$\pm$0.8  &  171$\pm$3   &   273$\pm$3                 \\
bkg model [d$^{-1}$kg$^{-1}$]    &	 11.5      &   35.4     &     22.4         &          196         &      266            \\ \hline
prompt $\mu$-induced &	    35\%         &      36\%                             &      24\%                &     33\%                         &  65\%                        \\
metastable Ge states &	     0.4\%        &    0.3\%                       &        0.3\%           &   11\%                    &       2\%                 \\
$\mu$-induced n concrete &	3\%             &             3\%              &   0.2\%                &       0.3\%                &      0.2\%                  \\
cosmogenic  &	    2\%         & 3\%                          &    0.2\%               & 21\%                      &        1\%                \\
$^{210}$Pb shield  &	          0.8\%   &     1\%                      &     3\%              &     3\%                  &  13\%                      \\
$^{210}$Pb within cryostat &	   52\%          &   43\%                        &  71\%                 &    30\%                   &    15\%                    \\
other cont. within cryostat   &	   1\%          & 1\%                          &     1\%              &   1\%                    &       2\%                   \\ 
airborne Rn  &	   6\%          &     13\%                      &    0.5\%               &    0.6\%                   &   2\%                    \\\hline

\end{tabular}
\end{footnotesize}
\caption{Overview of the measured count rates in comparison to the MC background model within various energy ranges for C1, C2 and C3. The relative contribution of the different background sources is also shown in each energy range. If available, reactor OFF data is used for the comparison to the MC. This is not always the case for the highE range. 
($^{(*)}$data collected in optimization phase after RUN-1) 
\label{tab:bkgmodeltable}}
\end{center}
\end{table*}

\section{Summary and conclusions}
\label{chapter7}
 
An excellent suppression and understanding of the background are prerequisites for a CE$\nu$NS detection at reactor-site. 
The small overburden, the limitations of a reactor environment compared to a normal laboratory, as well as the reactor as potential additional background source pose challenges on the design of the shield. MC simulations are a valuable tool to understand all background contributions and to describe the measured background of the fully shielded experiment. The detailed characterization and understanding of the CONUS spectrometers enable an accurate MC simulation, especially regarding the creation of slow pulses which contribute to the background at low energies.  

To suppress external radiation, the CONUS shield consists of 25\,cm of Pb. The shallow overburden of effective 24\,m~w.e. at the experimental site makes the $\mu$-induced secondaries created in the Pb shield layers the dominant background source. The background is reproduced excellently with the help of MC simulations, in which muons are propagated through the shield layers. In this way, it can be derived that below 50\,keV$_{ee}$ this background consists mostly of recoils created by $\mu$-induced neutrons.
The background is successfully suppressed by the $\mu$-veto with an estimated background reduction efficiency of $\sim$97\%. With the applied $\mu$-veto, about (30-50)\% of the background below 1\,keV$_{ee}$ still corresponds to prompt $\mu$-induced signals.

Next to the $\mu$-induced background the remaining continuum is mainly made up out of the decays of $^{210}$Pb (T$_{1/2}$=22.3\,a), often found in Pb bricks, and its daughter nuclides. For the innermost layer of the CONUS shield a mean activity of $<$1.7\,Bq~kg$^{-1}$ was achieved by a careful material selection. 
 
Next to the shield layers, $^{210}$Pb decays are also observed within the cryostat end caps. These decays must occur on the surfaces of the Ge diodes and the Cu holders. 
They result in a line at 46.5\,keV as well as contributions to the continuum. 
Overall, about 20-50\% of the background below 1\,keV$_{ee}$ is assigned to this component according to the MC-based decomposition.
Other contaminations of materials inside the detector end caps are considered negligible, as there are nearly no indications of the corresponding background lines from natural radioactivity observed in the energy range available.
This is also due to the dedicated material screening and selection campaign before the detectors were manufactured. 

Most of the other observed remaining background lines are decays of metastable Ge states, that are created in the capture of $\mu$-induced neutrons, as well as cosmogenic-induced isotopes.
From the MC simulation, the contribution to these decays within the ROIs is shown to be small. Nevertheless, the half-life of some of these cosmogenic isotopes, created in spallation reactions at Earth's surface in Ge and Cu, is of the order of typical physics data collection runs. However, the detailed MC simulation of these contributions showed that in the ROIs changes of less than 0.1-0.2\,d$^{-1}$kg$^{-1}$ in [0.3,1]\,keV$_{ee}$ and less than 0.3-0.5\,d$^{-1}$kg$^{-1}$ in [2,8]\,keV$_{ee}$ in the period of one year are expected at maximum. For the current analyses they are considered negligible, but they can also be corrected using the MC spectra if necessary. By studying this decaying background contribution and combining it with the exposure history of the CONUS detectors, it was also possible to derive production rates for $^{68}$Ge, $^{75}$Zn and $^{57}$Co from the CONUS data.

Airborne Rn, varying over time, is kept away from the detectors by constantly flushing the detector chamber with Rn-free air from breathing air bottles that have been stored long enough for the Rn to decay away. In the middle of the outage of RUN-2, the Rn background was fully suppressed. Before, depending on the detector and run, minor Rn contributions (indicated by the 351.9\,keV line of $^{214}$Pb) were still present and are taken into account in the MC background model.

The reactor-correlated background at the exact location of the experiment has been thoroughly characterized already before the start of the experiment. The background needs to be understood in detail as it can potentially create differences in the collected reactor ON and OFF data sets and mimic CE$\nu$NS signals. The measured neutron fluence rate was propagated through the shield in the MC simulation, showing that this background is negligible for the physics analyses. Inhomegeneities in the neutron fluence rate within A-408 were found, showing that characterization measurements of this type at the exact location of the experiment are important.  

All MC spectra are combined individually to a background model for each detector. A Kolmogorow-Smirnow-test was used to confirm that the MC simulations provide an accurate description of the reactor OFF data.
The background model is used as input for the likelihood analyses looking for CE$\nu$NS or BSM physics. To completely describe the background spectrum in the ROI for CE$\nu$NS, additionally, a parameterization of the noise component is required. The noise starts to rise below $\sim$400\,eV$_{ee}$. For the BSM models at higher energies extending over a larger energy range of [2,8]\,keV$_{ee}$, small systematic uncertainties on the background shape are included in the likelihood fit routine via pull terms. They arise from uncertainties on the production rates of the cosmogenic isotopes reported in literature as well as on the thickness of the detector's passivation layer influencing the shape of the $^{210}$Pb spectrum from the decay on the surface of the Ge diodes.

All in all, the compact CONUS shield is able to suppress the background at reactor-site at shallow depth by more than four orders of magnitude. A full decomposition of the background spectra in the various contributions described by MC simulations over the full lowE and highE range was achieved. All relevant components as well as detector properties and efficiencies such as slow pulses are included in the MC simulation. To our knowledge, this is the first time, that this was achieved for a Ge-based experiment operated at reactor-site. In this way, the significance of the different background contributions could be examined.  

In the upcoming CONUS data collection, improvements in the background suppression will be possible by including the knowledge gained from the pulse shape of the measured signals. It will be possible to suppress slow pulses and reduce as well as flatten the background in the ROI. As the currently used DAQ is not able to provide this information, a new system has been installed. Accompanying the pulse shape analysis with electric field calculations and the interaction points of the particles in the MC will further enhance our understanding of the background data of the CONUS experiment.

\section{Acknowledgements}
\label{sec:acknowlegments}
We are grateful to the late Dr. C. Schlosser for providing us with the Freiburg Minster lead. We thank Dr. M. W\'{o}jcik (Jagiellonian University, Krakow, Poland) for the determination of the $^{210}$Pb activity of some of the bricks within the CONUS shield and Dr. M. Laubenstein (LNGS, Assergi, Italy) for providing material screening measurements of various detector components. 
Moreover, we express our gratitude to the construction department and mechanical workshops of MPIK regarding the design and setup of the CONUS shield, especially M. Rei{\ss}felder, A. Schwarz and T. Apfel.   
We thank N. Nagel (KBR) for his effort to provide a clean environment during the setup of the shield and afterwards. We are indebted to T. Schr\"oder (KBR) and S. Winter (KBR) for their continuous support in the exchange of the breathing air bottles for Rn mitigation.

\bibliographystyle{unsrt}

\end{document}